\documentclass[%
 aip,
 amsmath,amssymb,
 reprint,%
 floatfix,
]{revtex4-1}

\usepackage{graphicx}
\usepackage{dcolumn}
\usepackage{bm}

\usepackage[utf8]{inputenc}
\usepackage[T1]{fontenc}
\usepackage{mathptmx}
\usepackage{xcolor}

\begin{document}

\preprint{AIP/123-QED}

\title[Nonequilibrium dynamics in polymer melts under orthogonal interrupted shear]{Probing the nonequilibrium dynamics of stress, orientation and entanglements in polymer melts with orthogonal interrupted shear simulations.}

\author{Marco A. Galvani Cunha}
\email{mgalvani@sas.upenn.edu}
\affiliation{ 
Department of Physics \& Astronomy, University of Pennsylvania
}%

\author{Peter D. Olmsted}%
\affiliation{ 
Department of Physics and Institute for Soft Matter Synthesis \& Metrology, Georgetown University
}

\author{Mark O. Robbins}
\affiliation{%
Department of Physics \& Astronomy, Johns Hopkins University
}%

\date{\today}

\begin{abstract}

Both entangled and unentangled polymer melts exhibit stress overshoots when subject to shearing flow. The size of the overshoot depends on the applied shear rate and is related to relaxation mechanisms such as reptation, chain stretch and convective constraint release. Previous experimental work shows that melts subjected to interrupted shear flows exhibit a smaller overshoot when sheared after partial relaxation. This has been shown to be consistent with predictions by constitutive models. Here, we report molecular dynamics simulations of interrupted shear of polymer melts where the shear flow after the relaxation stage is orthogonal to the original applied flow. We observe that, for a given relaxation time, the size of the stress overshoot under orthogonal interrupted shear is larger than observed during parallel interrupted shear, which is not captured by constitutive models. Differences in maxima are also observed for overshoots in the first normal stress and chain end-to-end distance. We also show that measurements of the average number of entanglements per chain and average orientation at different scales along the chain are affected by the change in shear direction, leading to non-monotonic relaxation of the off-diagonal components of orientation and an appearance of a 'double peak' in the average number of entanglements during the transient. We propose that such complex behavior of entanglements is responsible for the increase in the overshoots of stress components, and that models of the dynamics of entanglements might be improved upon by considering a tensorial measurement of entanglements that can be coupled to orientation.

\end{abstract}

\maketitle

\section{\label{sec:Intro}Introduction}

The concept of entanglements is fundamental to the theory of polymer dynamics\cite{de_gennes_reptation_1971,doi_theory_1988}. In equilibrium, tube theories relate the timescales that govern the dynamics of individual polymer chains to the rheological properties of the melt by scaling relations that involve the number of entanglements per chain. However, it is not clear how this picture can be fully extended to the nonequilibrium regime. Changes in the entanglement structure are one of several relaxation mechanisms that play a role in the complex nonlinear viscoelastic behavior of entangled linear polymer melts under shear\cite{marrucci_dynamics_1996,milner_microscopic_2001}. Dynamic equations for disentanglement have been proposed\cite{ianniruberto_convective_2014}, but one interesting effect that has been overlooked is the coupling of entanglements to orientation and how this affects the relaxation mechanisms. Entanglements are usually built into constitutive models through the number of entanglements per chain, which is a scalar quantity. However, in melts that have been highly oriented by flow the interaction of any chain with its neighbors becomes anisotropic. Neglecting this property can lead to problems when using constitutive models to calculate the change in components of stress under flow, when starting from an out of equilibrium state. In this paper, we simulate an orthogonal flow protocol to study the nonlinear rheology of startup shear, starting from states that are partially relaxed after being driven to steady state of shear, and compare the simulation results to predictions by widely used constitutive equations. We also analyze the evolution of the number of entanglements per chain and the orientation at multiple scales along the chain during the transient period of startup.

Several models for the nonlinear viscoelasticity of polymer melts have been proposed. The original tube model of Doi and Edwards\cite{doi_theory_1988} relates the polymer stress to the average orientation of the tube, and integrates over the whole deformation history of the melt. More recent models incorporated other relaxation mechanisms, such as contour-length fluctuations, chain stretch and convective constraint release (CCR)\cite{milner_reptation_1998,marrucci_dynamics_1996,milner_microscopic_2001,ianniruberto_convective_2014,ianniruberto_simple_2001}. Some of these are the DEMG model\cite{marrucci1988fast}, which includes the effects of chain stretch on the simpler Doi-Edwards theory; the MLD model\cite{mead_molecular_1998}, which incorporates CCR; and the GLaMM model\cite{graham_microscopic_2003}, which starts from a stochastic differential equation describing the microscopic dynamics of a polymer chain that can be used to derive an equation for the stress. Another approach not based on tube theory is the slip-link model\cite{hua_segment_1998}. One of the more successful simple models proposed is the Rolie-Poly model\cite{likhtman_simple_2003}, which includes reptation, chain stretch and CCR as relaxation mechanisms, and is a single-mode version of the GLaMM model\cite{graham_microscopic_2003}. It has been shown to describe, at least semi-quantitatively, several observations of nonlinear phenomena under many flow conditions, in both experiment and simulation \cite{likhtman_simple_2003,cao_simulating_2015}.

One of the hallmarks of nonlinear rheology of polymer melts is the existence of stress overshoots during startup shear under strong flows\cite{boukany_universal_2009,dealy_structure_2018}, particularly with Rouse-Weissenberg numbers $Wi_R > 1$. Overshoots have been observed extensively in both experiments and simulations\cite{wang_new_2013,robertson_reentanglement_2004,cao_simulating_2015,nafar_sefiddashti_steady_2016,nafar_sefiddashti_evaluation_2017,nafar_sefiddashti_individual_2019,nafar_sefiddashti_elucidating_2019,anwar_nonlinear_2019}, and connecting the overshoot in the stress components to changes in molecular properties has been a subject of debate in the literature\cite{mohagheghi_elucidating_2016,nafar_sefiddashti_individual_2015}. The strain at the peak of the viscosity $\eta$ varies from $\gamma^*_{\eta} \approx$ 1 up to $\gamma^*_{\eta} \approx$ 10. The first normal stress difference $N_1$ also goes through an overshoot, and its peak happens at a strain $\gamma^*_{N_1} \approx 2\gamma^*_{\eta}$. Simulations find the same qualitative features, and also show that the overshoot in the stresses is mirrored by the components of the chain orientation tensor for low and moderate shear rates \cite{jeong_effect_2017,jeong_molecular_2017,cao_simulating_2015}. This is usually interpreted as a validation of the stress optical rule, which states that information about the stress is encoded in the orientation and can be correlated to birefringence. Simulations also show an overshoot in the stretch and an undershoot in the number of entanglements per chain at very high rates.

One interesting protocol for investigating nonlinear rheology of polymer melts is interrupted startup shear\cite{stratton_stress_1973,tsang_use_1981,robertson_reentanglement_2004}. Experiments have shown that when one interrupts shear after reaching steady-state and resumes it after a short waiting time, the observed transient overshoot is smaller during the second stage of shear. This has been attributed to changes in the entanglement structure of the melt during the first shear\cite{roy_reentanglement_2013}, but reductions in the overshoot during the second shear have also been observed for unentangled melts\cite{santangelo_interrupted_2001}. Associated changes on the first normal stress overshoot have also been observed. Interestingly, while the original overshoot for unentangled systems is recovered for waiting times similar to the longest equilibrium relaxation time (the Rouse time), for entangled systems it is only recovered on timescales longer than several disengagement times. 

While this protocol is useful for studying the effects of relaxation on the size of the overshoot, it is challenging to perform in both experiments and simulations due to the unusual changes in the flow over time. However, it allows for a new type of experiment that so far has not been realized: restarting shear with a different flow direction after relaxation. Such a setup can probe the changes in relaxation due to starting shear from a configuration far from equilibrium. For an aligned chain, the local molecular environment is different in the directions along the backbone and in the plane orthogonal to the backbone, and this anisotropy in the entanglement structure might lead to a different stress response. A related concept is orthogonal superposition rheometry, where a fluid in steady-state of shear is subject to oscillatory shear in a direction orthogonal to the steady state flow\cite{wong_orthogonal_1989,kwon_remarks_1993,moghimi_orthogonal_2019,jiamin_modeling_2021,vermant_orthogonal_1998,mead_small_2013,metri_brownian_2019,ianniruberto_superposition_2015,kim_superposition_2013}. A similar set of experiments was conducted by Hurlimann and Kraft \cite{hurlimann_multidirectional_1991,kraft_shear_1997,kraft_untersuchungen_1996} in a series of papers where they measured the rheological response of low-density polyethylene melts under a sudden change in shear direction using a multidirectional rheometer. Their experiments are essentially orthogonal interrupted shear experiments with no waiting time between the end of the first stage of shear and start of the second stage. Their focus was on the changes observed in the overshoot in the viscosity when the first stage of shear was interrupted at different pre-strains. For pre-strains smaller than the strain needed to reach the peak of the overshoot, little change was observed on the size of the overshoot during the orthogonal stage of shear. However, for large pre-strains (larger than the strain needed to reach steady state) a significant reduction on the overshoot during the orthogonal shear stage was observed. Jeyaseelan and Giacomin \cite{jeyaseelan_polymer_1995} showed that transient network theory predictions were in reasonable agreement with these experiments, but the absence of strong overshoots in the stress for the analyzed rates makes agreement during the transient regime easier.

Constitutive equations based on tube theory can capture the qualitative features of interrupted shear experiments. The Rolie-Poly model can predict not only the monotonic recovery of the overshoot as a function of waiting time when the first shear is stopped after steady state, but also the non-monotonic recovery seen if one stops the first stage of shear during the overshoot\cite{graham_comment_2013,holroyd_analytic_2017}. The Doi-Edwards model, which only includes relaxation due to orientation of the chains, is also able to model the recovery of the overshoot, suggesting that while disentanglement might contribute quantitatively to the recovery of the overshoot it is not the only mechanism responsible for it\cite{ianniruberto_repeated_2014}.

In this paper we describe non-equilibrium molecular dynamics simulations of interrupted startup shear flow. We perform simulations in which the second shear is in the same direction as the original, or in a plane orthogonal to the plane of shear during the first startup. The stress overshoot is significantly larger for a given waiting time for an orthogonal second shear. We also show measurements of the normal stresses as well as molecular scale quantities such as the end-to-end distance and orientation during both stages of startup shear. All of the relevant measures show corresponding overshoots and undershoots before reaching steady-state, though the time at which they occur are not the same for all quantities. We contrast these measurements to predictions by constitutive equations based on tube theory and show that available models are not capable of capturing this behavior even qualitatively.

\section{\label{sec:Methods}Methods}

\subsection{\label{subsec:MD}Molecular Dynamics Simulations}

Molecular dynamics simulations were carried with the package LAMMPS \cite{lammps} and used the Kremer-Grest bead-spring model for linear homopolymers\cite{kremer_dynamics_1990}. Monomers were modeled as beads with mass $m$ interacting via a purely repulsive truncated Lennard-Jones (LJ) potential
\begin{equation}
    U_{LJ} = 4u_0\left[\left(\frac{a}{r}\right)^{12}-\left(\frac{a}{r}\right)^6+1\right] \text{for $r < r_c = 2^{1/6}a$}.
\end{equation}
All simulation results will be in terms of LJ units (distance $a$, energy $u_0$, mass $m$ and time $\tau = a\sqrt{m/u_0}$). The simulated melts have a total of 184,000 beads, and each melt comprises polymers with length $N =$ 40 or 500 beads. The polymers are linear chains of beads connected by covalent bonds represented by the FENE potential
\begin{equation}
    U_{FENE} = 1.5KR_0^2\ln\left[1-(r/R_0)^2\right]
\end{equation}
where $K = 30u_0/a^2$ and $R_0 = 1.5a$. All simulations were performed at a density of $\rho = 0.85$.

Table \ref{tab:sys} lists parameters for the systems studied. These parameters have been measured extensively in the literature \cite{g_cunha_effect_2020,cao_simulating_2015,auhl_equilibration_2003}. Recently Svaneborg and Everaers have published values of these parameters for many different stiffnesses \cite{svaneborg_characteristic_2020}, and values in the table are equal to the values quoted in their paper when these are readily available. All values quoted were consistent with direct measurements from our simulations. For the timescales, we use the values in the literature for the entanglement Rouse time $\tau_e$ and obtain the chain Rouse time $\tau_R$ and disentanglement time $\tau_d$ from scaling relations. The scaling relation for the Rouse time is the standard $\tau_R = \tau_e Z^2$, while for $\tau_d$ we use the scaling relationship by Likhtman et al.\cite{likhtman_quantitative_2002} which accounts for contour length fluctuations:
\begin{equation}
    \tau_d/\tau_R = 3Z\left(1-\frac{3.38}{Z^{1/2}}+\frac{4.17}{Z}-\frac{1.55}{Z^{3/2}}\right)
    \label{eq:taud}
\end{equation}
where $Z = N/N_e$ and $N_e$ is the entanglement length.

\begin{table*}
\caption{Parameters of the different types of systems used for MD simulations. For each number of beads per chain $N$ we measure the end-to-end length$\langle R_{ee}^2\rangle^{1/2}$, Kuhn length $l_k$, entanglement length $N_e$, number of entanglements $Z$ and the entanglement, Rouse and disentanglement times. Values for the entanglement times are obtained from Ref. \cite{svaneborg_characteristic_2020}.}
    \begin{ruledtabular}
    \begin{tabular}{ccccccccc}
        $N$ & $\langle R_{ee}^2\rangle^{1/2}/a$ & $l_k/a$ & $N_e$ & $Z \equiv N/N_e$ & $\tau_e/\tau$ & $\tau_R/\tau$ & $\tau_d/\tau$ \\
        \hline
        40 & 8 $\pm$ 1 & 1.8 & 86 $\pm$ 7 & $<$1 & -- & $2.4 \times 10^3$ & --\\
        500 & 29 $\pm$ 1 & 1.8 & 86 $\pm$ 7 & 6 $\pm$ 1 & $1.07 \times 10^4$ & $3.7 \times 10^5$ & $1.2 \times 10^6$\\
    \end{tabular}
    \end{ruledtabular}
    \label{tab:sys}
\end{table*}

For all simulations, the equations of motion were integrated with a timestep $\delta t$ = 0.01$\tau$, and temperature was kept at $T = 1.0u_0$. Equilibrated melts were generated using the standard double-bridging algorithm \cite{auhl_equilibration_2003}. Temperature during equilibration was controlled by a Langevin thermostat with a time constant of 10$\tau$.

\subsection{Shear Protocols}

The simulation protocol involved two stages of simple shear with a stage of relaxation in between. Figure \ref{fig:flows} shows the shear plane and rate used during each stage of the simulations. In the first stage the simulation box is sheared in the $xy$-plane at a desired rate. After the simulation has reached steady state, we stop the flow in a configuration where the simulation box is orthorhombic. The melt is allowed to relax for a certain time $t_w$ and then sheared again. During the second stage of shear we deform the box either in the $xy$ or $yz$ plane (where the first axis refers to the flow direction and the second refers to the gradient direction), leading to different transient stress responses. The rates used for shearing each system were chosen to produce Rouse Weissenberg numbers $Wi_R = \dot{\gamma}\tau_R$ ranging from 1 to 250. In this flow regime, nonlinear viscoelastic effects become relevant and the transient overshoot in the viscosity is very prominent \cite{cao_simulating_2015,jeong_effect_2017,baig_flow_2010,nafar_sefiddashti_individual_2019}.

During the relaxation stage, temperature was controlled by a Langevin thermostat with a time constant of 10$\tau$\cite{tuckerman_statistical_2010}. The temperature is obtained from the velocities of atoms after subtracting an instantaneous linear velocity profile that is calculated by binning the system in the $y$ direction and calculating the average $x$ velocity in each bin. This is done so that the remaining velocity profile after the first stage of shear does not affect the temperature at the beginning of the relaxation stage. Other methods for calculating temperature such as accounting only for the velocity component in the vorticity direction were also used and resulted in no measurable difference in the measured temperature and stresses.

Both stages of shear were simulated by integrating the SLLOD equations of motion and deforming the simulation box with the desired shear rate, and temperature was controlled with a Nose-Hoover thermostat with a time constant of 10$\tau$\cite{todd_nonequilibrium_2017}. In our simulations, the flow homogeneity was ensured by directly probing the linear velocity profile across the system, and peculiar velocities were defined by subtracting the profile at a particle's position from its velocity. During these stages the thermostat was biased by subtracting a linear velocity profile consistent with the simulation box deformation rate before thermostatting, and then reinstating the linear profile after the thermostatting was performed on the peculiar velocities.

Since the SLLOD equations of motion simulate a system under homogeneous flow, inhomogeneities such as shear banding are not seen in our simulations\cite{todd_nonequilibrium_2017}. Other methods can be used to probe inhomogeneous flows. Recent Dissipative Particle Dynamics simulations with Lees-Edwards boundary conditions have shown that shear banding can be observed during startup shear of entangled polymer melts\cite{boudaghi-khajehnobar_effects_2020,mohagheghi_molecularly_2016}. In that case, the melt separates into fast and slow bands where stress overshoots are still seen, but the values of the normal stress maxima differ from one band to the other.

The different types of thermostatting can lead to artifacts when calculating the temperature immediately after the flow is turned off or on. This is particularly important for the systems with small chains, where the relevant timescales of the dynamics are shorter and short time stress data can be affected by temperature fluctuations. In order to avoid this problem, the shearing stages were initialized with a velocity profile consistent with the desired shear rate, and peculiar velocities were adjusted to match the desired temperature. Other methods such as thermostatting only the direction perpendicular to the flow plane and quickly ramping the shear rate from zero to the desired value at the start of shear resulted in no significant change in the measured values of the stress in all cases tested.

\subsection{Analysis}

Rheological data is obtained during the transient regime of startup shear and interrupted shear. Due to the noisy nature of stress measurements at individual timesteps, we report stress data for single simulations that have been filtered using a Savitzky-Golay filter \cite{savitzky_smoothing_1964}. This filtering method works by fitting polynomials to sequential subsets of data points, effectively averaging data over a moving window without distorting the underlying signal.

Individual chain statistics such as length and orientation along different axes at different scales can be obtained by taking suitable averages over the positions of the monomers. Primitive path information can be obtained by one of several methods published in the literature\cite{hoy_topological_2009,everaers_rheology_2004,sukumaran_identifying_2005}, usually by an algorithm that shrinks the backbone of the chains as short as possible without allowing chains to pass through each other. 

Even in equilibrium, measurements of the number of entanglements by different estimators can vary; equivalently, one must determine the \textit{entanglement length} $N_e$, \textit{i.e.} the number of bonds per entanglement. Hoy and Kr\"oger \cite{hoy_topological_2009} showed how different ways of calculating the entanglement length lead to different values. One possible measure, based on random walk statistics,  is the rheological entanglement length $N_e^{rheol} \equiv N_b\langle R^2\rangle/L_{pp}^2$, where $N_b$ is the number of bonds, $\langle R^2\rangle$ is the average squared end-to-end distance and $L_{pp}$ is the primitive path length of a chain. One can also calculate a topological entanglement length, given by $N_e^{topol} = N_b/Z_k$ where $Z_k$ is the number of topological constraints (points of contact between chains as obtained by an algorithm that reduces chains to their primitive paths) on a chain. It is usually found that the rheological entanglement length is larger than the topological entanglement length by a factor of around two. To avoid confusion we will use $N_e$ to refer to the rheological entanglement length in equilibrium, and this is the definition used for values in table \ref{tab:sys} and in scaling relations. Since this method does not work away from equilibrium, we use the Z1 code\cite{kroger_shortest_2005} to obtain primitive path statistics as well as information about contact between the primitive paths of different chains. The Z1 code works by shrinking the polymer chains to their primitive paths through a series of geometrical moves and identifying points where a move would force two chains to come in contact or cross. We call these contact points topological constraints (TCs) and analyze the evolution of the average number of TCs per chain $\langle Z_k\rangle$ under flow. Other methods such as Primitive Path Analysis\cite{sukumaran_identifying_2005} can generate similar information, and direct comparison between TC statistics generated by different algorithms in aligned melts is the subject of undergoing research. 

\subsection{\label{subsec:TM}Tube Model Predictions}

We use the Rolie-Poly model as well as a differential version of the Doi-Edwards model \cite{larson_constitutive_1984} to calculate the stress tensor during both stages of startup shear as well as during relaxation. The differential version of the Doi-Edwards model is
\begin{equation}
\frac{d\bm{\sigma}}{dt} = \bm{\kappa\cdot\sigma + \sigma\cdot\kappa^T}-\frac{1}{\tau_d}(\bm{\sigma-I})- \frac{2}{3}\bm{\kappa : \sigma}\bm{\sigma},
\label{eq:DE}
\end{equation}
where $\bm{\sigma}$ is the polymer stress and $\bm{\kappa}$ is the velocity gradient tensor given by $\kappa_{ij} = \partial_j v_i$. The operator $d/dt$ is the material derivative. This differential approximation to the original integral Doi-Edwards model and is more suitable to numerical analysis. It has been shown to exhibit much of the same behavior as the integral version, except that it predicts a zero second normal stress difference. Like its integral counterpart, this model does not show an overshoot on the first normal stress difference, predicting a monotonically increasing $N_1$ that saturates at the steady-state value \cite{dealy_structure_2018}.

\begin{figure}
    \centering
    \includegraphics[width=\columnwidth]{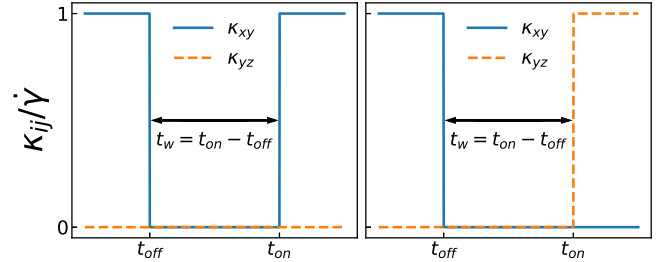}
    \caption[Components of velocity gradient tensor in interrupted startup shear.]{Components of the velocity gradient tensor for parallel interrupted shear (left) and orthogonal interrupted shear (right) simulations. The flow is turned off and on during the interrupted shear protocol at times  $t_\textrm{off}$ and $t_\textrm{on}$. The rates utilized always have the same magnitude in both stages of shear.}
    \label{fig:flows}
\end{figure}

The Rolie-Poly \cite{likhtman_simple_2003} equation is a single mode version of the GLaMM \cite{graham_microscopic_2003} model that includes contributions to relaxation from reptation, retraction and convective constraint release (CCR). The Rolie-Poly constitutive equation is
\begin{eqnarray}
\frac{d\bm{\sigma}}{dt} &=& \bm{\kappa\cdot\sigma + \sigma\cdot\kappa^T}-\frac{1}{\tau_d}(\bm{\sigma-I})\nonumber \\
 & &-\frac{2(1-\sqrt{3/tr\bm{\sigma}})}{\tau_R}\left(\bm{\sigma}+\beta\sqrt{\frac{3}{tr\bm{\sigma}}}(\bm{\sigma-I})\right),
 \label{eq:rp}
\end{eqnarray}
where the parameter $\beta$ sets the relative strength of the CCR relaxation. The disentanglement and Rouse times are related by Equation \ref{eq:taud}. To replicate the flow conditions in our repeated startup shear MD simulations, we use different velocity gradient tensors to represent the different stages of the MD simulation when calculating the polymer stress using the constitutive equations. The chosen $\bm{\kappa}$ controls which components of the stress tensor couple to flow in the advective terms of Equations \ref{eq:DE} and \ref{eq:rp}. The most general velocity gradient tensor needed for our Rolie-Poly calculations is given by 
\begin{equation}
    \bm{\kappa} =
    \begin{pmatrix}
      0 & \kappa_{xy} & 0\\
      0 & 0 & \kappa_{yz}\\
      0 & 0 & 0
    \end{pmatrix},
    \label{eq:kappa1}
\end{equation}
where we remind the reader that the first index defines the velocity direction $\bm{\textrm{v}}$ and the second index defines the gradient direction $\nabla$, \text{i.e.} $\kappa_{ab}\equiv\kappa_{\textrm{v}\nabla}$.
The two shear protocols are shown in figure \ref{fig:flows}. The first stage of shear always has nonzero $\kappa_{xy}$ ($\kappa_{yz}=0$), while the shear after relaxation can have either $\kappa_{xy}$ or $\kappa_{yz}$ as the nonzero component of the velocity gradient tensor. The choices $\kappa_{xy}$ or $\kappa_{yz}$ couple different components of the stress tensor to the flow during the second stage of shear while the uncoupled components continue to relax.

\section{\label{sec:Res}Results}

For both the parallel and orthogonal interrupted shear protocols, the viscosity $\eta$, normal stress differences $N_1$ and $N_2$ and chain statistics were measured for different values of shear rate $\dot{\gamma}$, waiting time $t_w$ between the two shear stages and parameters in table \ref{tab:sys}. We contrast the transient stresses during the second shear to the transient stresses during the first shear and show that the main features apparent in the transient period vary monotonically with waiting time $t_w$. The simulation data shows that the magnitude of the transient peak in all relevant properties is larger during the second shear when flow is resumed in a direction orthogonal to the previous shear. We then show that both the Doi-Edwards and Rolie-Poly models predict the same magnitude of the overshoot in viscosity regardless of the relative orientation of the flow during the second stage, contradicting the simulations.

\subsection{First stage of shear}

\begin{figure}
    \centering
    \includegraphics[width=\columnwidth]{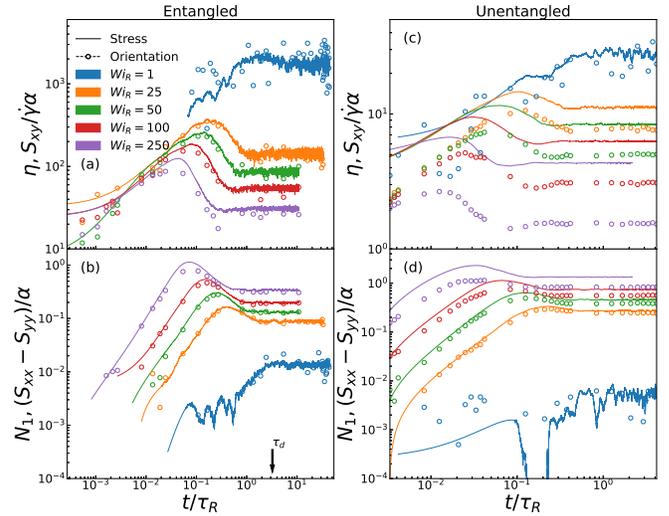}
    \caption[Stress and orientation during first startup.]{Viscosities and first normal stress differences (lines) at different Rouse Weissenberg numbers during the initial startup for entangled (a,b) and unentangled (c,d) fully-flexible systems. Components of the bond orientation tensor (circles) follow the stress closely at all but the highest rates for the entangled systems, satisfying the stress-optical rule (Equation \ref{eq:sor}). For the unentangled melt orientation and stress data only agree for the diagonal components of the stress.}
    \label{fig:startup}
\end{figure}

The measured values of $\eta$ and $N_1$ during the first stage of startup shear are summarized in Figure \ref{fig:startup}. Also plotted are components of the bond orientation tensor $S_{ij} = \langle u_{i}u_{j}\rangle$, where $\bm{u}$ is the bond vector connecting neighboring beads along the chain and $i$ and $j$ are cartesian indices. Measurements of the orientation are performed on a single snapshot of the simulation, leading to noisier data. We do not show data for $S_{yy}-S_{zz}$ or $N_2$, but their values are always negative and very small in magnitude, with an undershoot consistent with the observed stress behavior. Under startup shear, according to the stress-optical rule the stress components are proportional to the components of the bond orientation tensor,
\begin{equation}
    \sigma_{ij} = \frac{1}{\alpha}S_{ij},
    \label{eq:sor}
\end{equation}
where $\alpha$ is the stress-optical coefficient. For the fully-flexible FENE model, Cao and Likhtman \cite{cao_simulating_2015} showed that $\alpha = 0.32a^3/u_0$. For the entangled system, orientation and stress components track each other closely with minor deviations at the highest Weissenberg numbers. This can be attributed to the higher degrees of chain stretch and disentanglement induced at these stronger flows. For more moderate shear rates, the stress overshoot seems to be completely accounted for by the change in orientation, as predicted by the stress-optical law. For unentangled systems the shear stress is significantly larger than orientation for all Weissenberg numbers larger than one. One possible explanation for this is that for unentangled systems the polymer contribution to the stress is small and therefore on the same order of magnitude as the contribution to the stress from unentangled interchain interactions.

Previous simulations of startup shear flow of entangled polymer melts have also observed weak undershoots on $\langle Z_k\rangle$, the number of topological constraints per chain as measured by the Z1 code, during startup. \cite{nafar_sefiddashti_elucidating_2019} 

The strain at the peak of overshoot or undershoot is different for the shear and normal stresses. In experiments and simulations of both entangled and unentangled melts the strain at the peak of the overshoot in $N_1$ is usually twice as large as the peak strain in $\eta$\cite{menezes_nonlinear_1982,santangelo_interrupted_2001,jeong_molecular_2017,nafar_sefiddashti_individual_2019}. This is consistent with our findings, and we also observe even larger strains at the peak for the chain end-to-end length.

\subsection{\label{subsec:2nd_shear_xy}Parallel interrupted shear}

\begin{figure}
    \centering
    \includegraphics[width=\columnwidth]{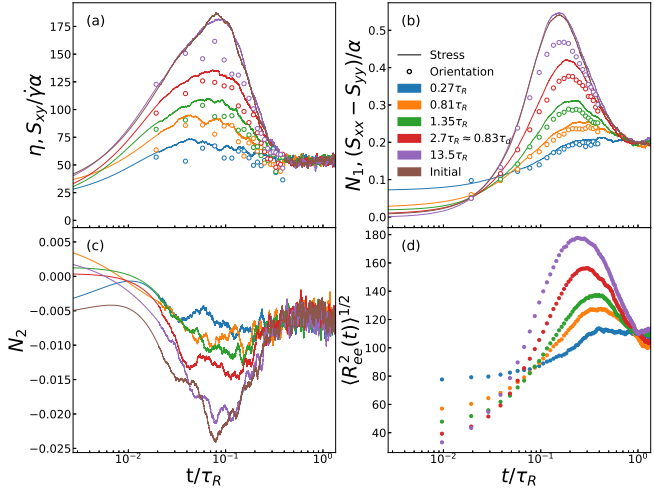}
    \caption[Stress and orientation during interrupted shear of an entangled melt.]{Stress and orientation components of the entangled fully-flexible melt during parallel interrupted shear simulations for different waiting times $t_w$. Plots show (a) viscosity, (b) first and (c) second normal stress differences and (d) end-to-end distance of chains. The curve labeled initial refers to the first stage of shear, and the other curves refer to the second stage of shear after a waiting time indicated by the label. Open circles in the plots for stress show the components of the bond orientation tensor $\bm{S} = \langle \bm{uu}\rangle$ properly normalized according to the stress-optical rule. Parameters: $N=500$, $Wi_R = 100$.}
    \label{fig:Wi100_long}
\end{figure}

Figure \ref{fig:Wi100_long} shows the viscosity, normal stress differences and end-to-end length measured during the first and second stages of shear for a melt of chains with N = 500 monomers at a Rouse Weissenberg number $Wi_R = 100$ and second shear in the same plane as the initial shear. Components of the bond orientation tensor during the overshoot are included for comparison. The orientation is correlated to the stress in a manner similar to the one observed during the startup from equilibrium, and the difference between stress and orientation is smaller for shorter waiting times. The size of the stress overshoot during startup shear increases monotonically with the waiting time $t_w$. The same behavior is observed for the first normal stress difference and end-to-end distance, as well as for the undershoot in the second normal stress difference.

\begin{figure}
    \centering
    \includegraphics[width=\columnwidth]{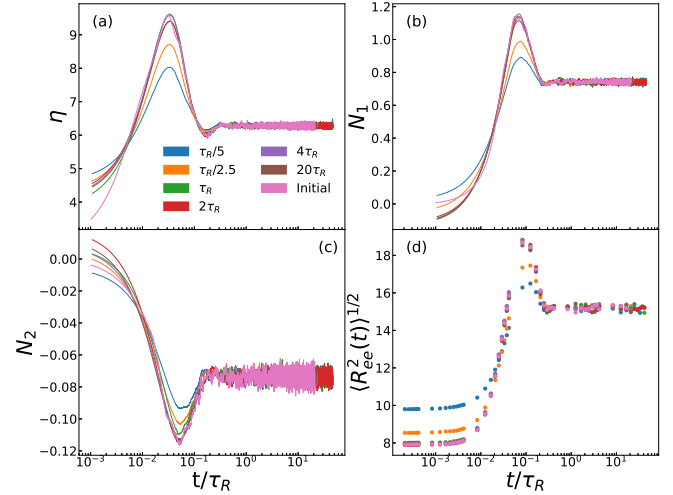}
    \caption[Stress and orientation during parallel interrupted shear of an unentangled melt.]{Stress components of the unentangled fully-flexible melt during parallel interrupted shear simulations. Plots show (a) viscosity, (b) first and (c) second normal stress differences and (d) end-to-end distance of chains. The curve labeled initial refers to the first stage of shear, and the other curves refer to the second stage of shear after a waiting time indicated by the label. Parameters: $N=40$, $Wi_R = 100$.}
    \label{fig:Wi100_short}
\end{figure}

Similar features can be observed when the simulations are repeated with a melt of short chains, as can be seen in fig. \ref{fig:Wi100_short}. The dependence of the size of the overshoot on the waiting time between stages of shear remains even in the absence of entanglements between chains. In the unentangled case full recovery of the overshoot occurs in the Rouse time, while for the entangled melt full recovery happens for waiting times longer than the disentanglement time.

Both sets of simulations show that the strain at the overshoot peak is about twice as large for $N_1$ as it is for $\eta$ and $N_2$. The strain at the peak of viscosity seems to be independent of the waiting time, while the strain at the peak of the first normal stress difference decreases for the entangled system, moving from a position very close to the onset of steady-state for short waiting times to a smaller value at the largest waiting times. The strain at the peak of the overshoot in the end-to-end distance of the chains also appears to decrease with increasing waiting time.

\begin{figure}
    \centering
    \includegraphics[width=\columnwidth]{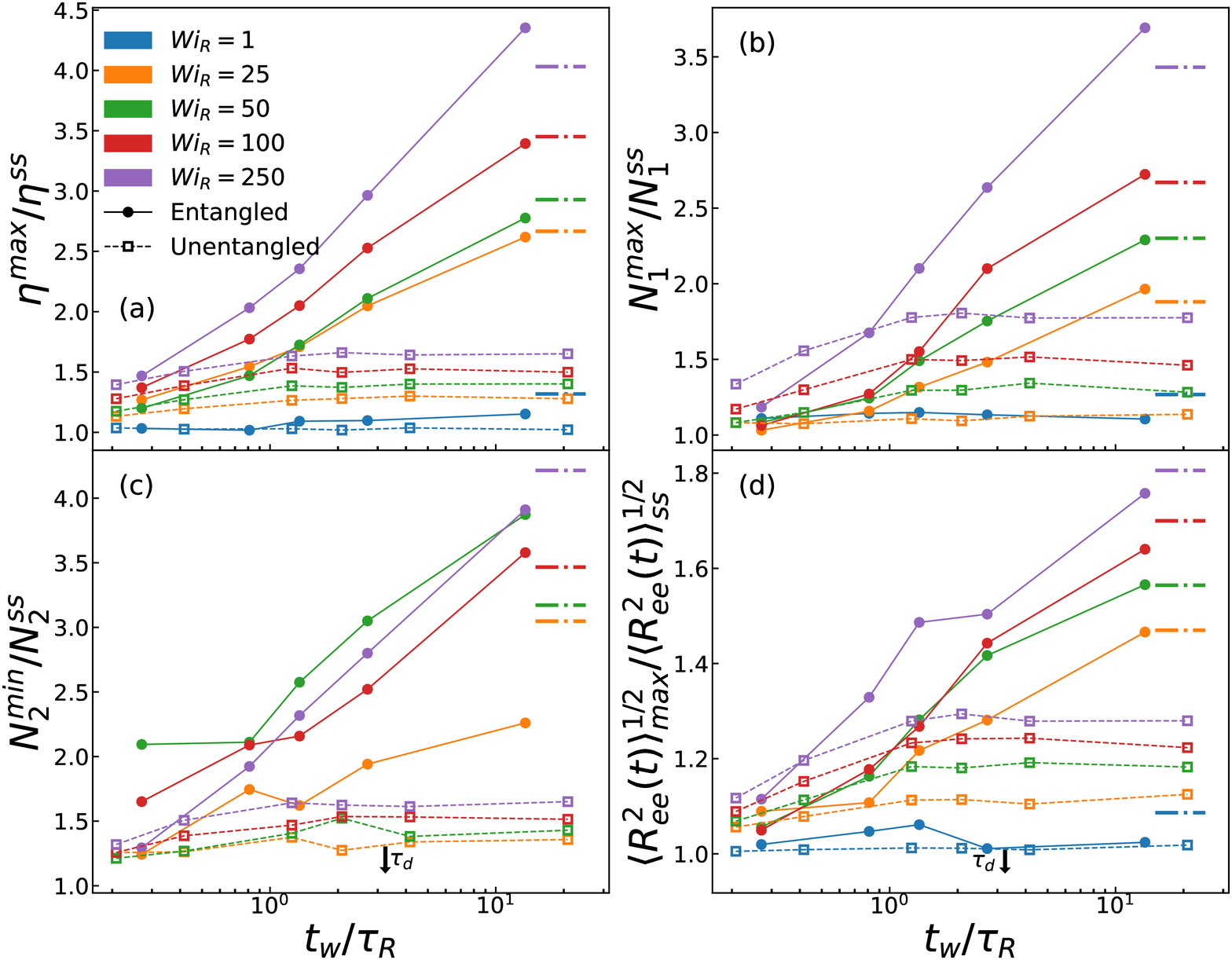}
    \caption[Relative size of overshoots as a function of waiting time.]{Relative size of overshoots and undershoots for entangled (filled circles) and unentangled (open squares) fully-flexible melts during parallel interrupted shear of fully flexible polymer melts as a function of waiting time. Peak of viscosity (a), $N_1$ (b) and end-to-end distance (d) overshoot normalized by steady-state value as a function of waiting time. (c) Minima of $N_2$ normalized by steady-state value.  The overshoots in the normal stresses for the unentangled case at the lowest Weissenberg number were too small to discern from noise and therefore are not plotted. All properties at the longest waiting time are statistically the same as startup from an equilibrium state. Dash-dotted lines represent the values for the initial startup at the different Weissenberg numbers for the entangled system. Arrow shows $\tau_d$ for the entangled system.}
    \label{fig:max_Wi_comp}
\end{figure}

Both the Doi-Edwards and Rolie-Poly models qualitatively capture the recovery of the overshoot in the viscosity\cite{graham_comment_2013,ianniruberto_repeated_2014}. Neither the Rolie-Poly or the differential Doi-Edwards model used here can predict a nonzero value for the second normal stress difference. The integral Doi-Edwards model predicts a value of $N_2 = -2N_1/7$, but neither form can predict an overshoot in $N_1$ \cite{larson_constitutive_1984}. Figure \ref{fig:max_Wi_comp} shows the values for the peak of the overshoot in the viscosity, $N_1$, $N_2$ and in the end-to-end distance obtained in the simulations as a function of waiting time. The maxima/minima for all properties for the unentangled system reach the terminal value (identical to the obtained when starting from equilibrium) after a waiting time around the Rouse time, which is the longest relaxation time for unentangled chains, while the entangled systems only reach the terminal value for the longest waiting times studied, which are over four times larger than the disentanglement time. This is consistent with trends observed in experiments on polystyrene melts \cite{santangelo_interrupted_2001}. A conservative estimate for the error bars on points in Fig. \ref{fig:max_Wi_comp} can be obtained by looking at the variation in the ratio $\eta^{max}/\eta^{ss}$ and other quantities during startup from several equilibrium configurations prepared independently. By simulating startup shear of 5 different equilibrium melt configurations, we estimate error bars to be of order $\pm 0.25$ in the viscosity and first normal stress ratios. Using the same method to estimate an error bar for the ratio of the second normal stress leads to an error bar of $\pm 0.75$. This shows that all melts at the longest waiting time are statistically equivalent to samples sheared from equilibrium.

Both the the entangled and unentangled systems exhibit a monotonic recovery of the overshoot, with corresponding changes to the bond orientation tensor. 

\begin{figure}
    \centering
    \includegraphics[width=\columnwidth]{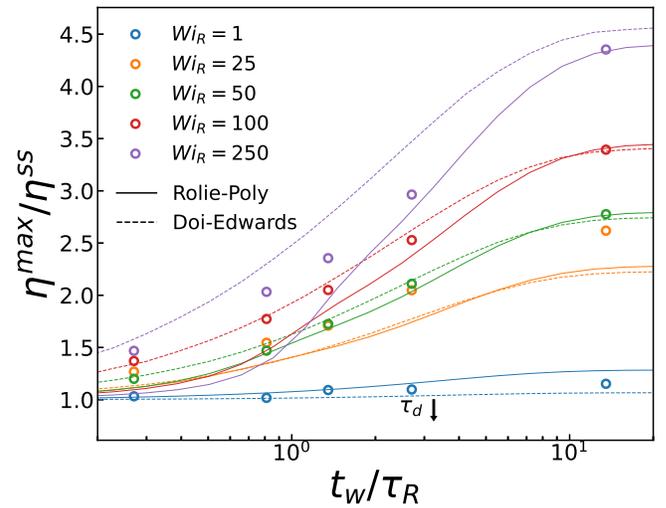}
    \caption[Comparison of constitutive equations to MD data for interrupted shear.]{Comparison of Doi-Edwards (dashed) and Rolie-Poly (solid) model predictions for relative size of the viscosity overshoot during the second stage of interrupted shear as a function of waiting time to simulation data for the fully-flexible entangled system.}
    \label{fig:data_RP_DE_comp}
\end{figure}

Figure \ref{fig:data_RP_DE_comp} compares predictions of the Doi-Edwards and Rolie-Poly models to the viscosity in the molecular dynamics simulations of the fully-flexible entangled system. The best Rolie-Poly model fit parameters were $Z = 9$ and $\beta = 1$, while the shown Doi-Edwards fit uses $Z = 3$. The Rolie-Poly value of $Z$ is consistent with $N_e = 60$, which is similar to the value used to fit the Likhtman-McLeish\cite{likhtman_quantitative_2002} and GLaMM\cite{graham_microscopic_2003} models to stress data by Cao and Likhtman\cite{cao_simulating_2015}. Although this value differs from the rheological entanglement length $N_e = 86$ reported in the literature, previous work has also shown that constitutive models using fit parameters derived from MD simulations do not necessarily correctly match the measured values of either the steady-state or the overshoot \cite{nafar_sefiddashti_elucidating_2019}. The value of $Z$ used in the Doi-Edwards fit is significantly smaller than the simulation and Rolie-Poly values. One possible explanation is the well-known excessive shear thinning predicted by the Doi-Edwards model. The ratio $\eta^{max}/\eta^{ss}$ grows very strongly with increasing $Wi_R$, so a small value of $Z$ is needed to fit the values measured in the simulations for our range of Weissenberg numbers. The Doi-Edwards model more accurately describes the data for shorter waiting times, while predicting smaller overshoots for longer waiting times. The best agreement with the Rolie-Poly model is for the longest waiting times. This might be due in part to our choice of using a single mode model, and better agreement for shorter waiting times can probably be obtained by adding a spectrum of shorter relaxation modes.

We also analyzed how the peak strain changes with waiting time. We see a systematic trend of a decrease by a factor of two for the strain at the peak of $N_1$ with increasing waiting time, while there is an increase by a factor of two for the viscosity peak strain. The latter is, however, hard to estimate since the shear stress curves tend to be flat near the peak, particularly for short waiting times, making it hard to measure the position of the peak precisely. Comparison with our chosen constitutive models is difficult, since the Doi-Edwards model predicts a strain at peak that does not vary with Weissenberg number and the Rolie-Poly model \cite{you_characteristic_2021} (as well as the more complicated GLaMM model\cite{anwar_nonlinear_2019}) has been shown to strongly overpredict $\gamma_max$ for flows in the regime $Wi_R > 10$. This also happens in our case, where our MD simulations show peak strains ranging from 2 to 11 for our different Weissenberg numbers, while Rolie-Poly predicts a peak strain of around 200 for our largest $Wi_R$. Rolie-Poly does predict a decrease of $\gamma_{max}$ for $N_1$  with increasing $t_w$, but also predicts a decrease for the peak strain of viscosity, which goes against what we observe in simulations.

\subsection{\label{subsec:2nd_shear_yz}Orthogonal interrupted shear}

We repeated the MD simulations detailed above with a second shear flow in a plane orthogonal to the first shear. Most of the same features remain, with overshoots and undershoots similar to those seen in the previous section. However, the size of the stress overshoot for the same waiting time is significantly larger when the direction of the second shear is changed.

In this orthogonal protocol, $y$ is now the flow direction and $z$ is the gradient direction (previously the vorticity). Hence both the flow and gradient directions are orthogonal. Later we will refer to this as ``fully orthogonal". Here the viscosity during the second stage of shear is given by $\eta = \sigma_{yz}/\dot{\gamma}$, and the normal stress differences are given by $N_1 = \sigma_{yy}-\sigma_{zz}$ and $N_2 = \sigma_{zz} - \sigma_{xx}$.

\begin{figure}
    \centering
    \includegraphics[width=\columnwidth]{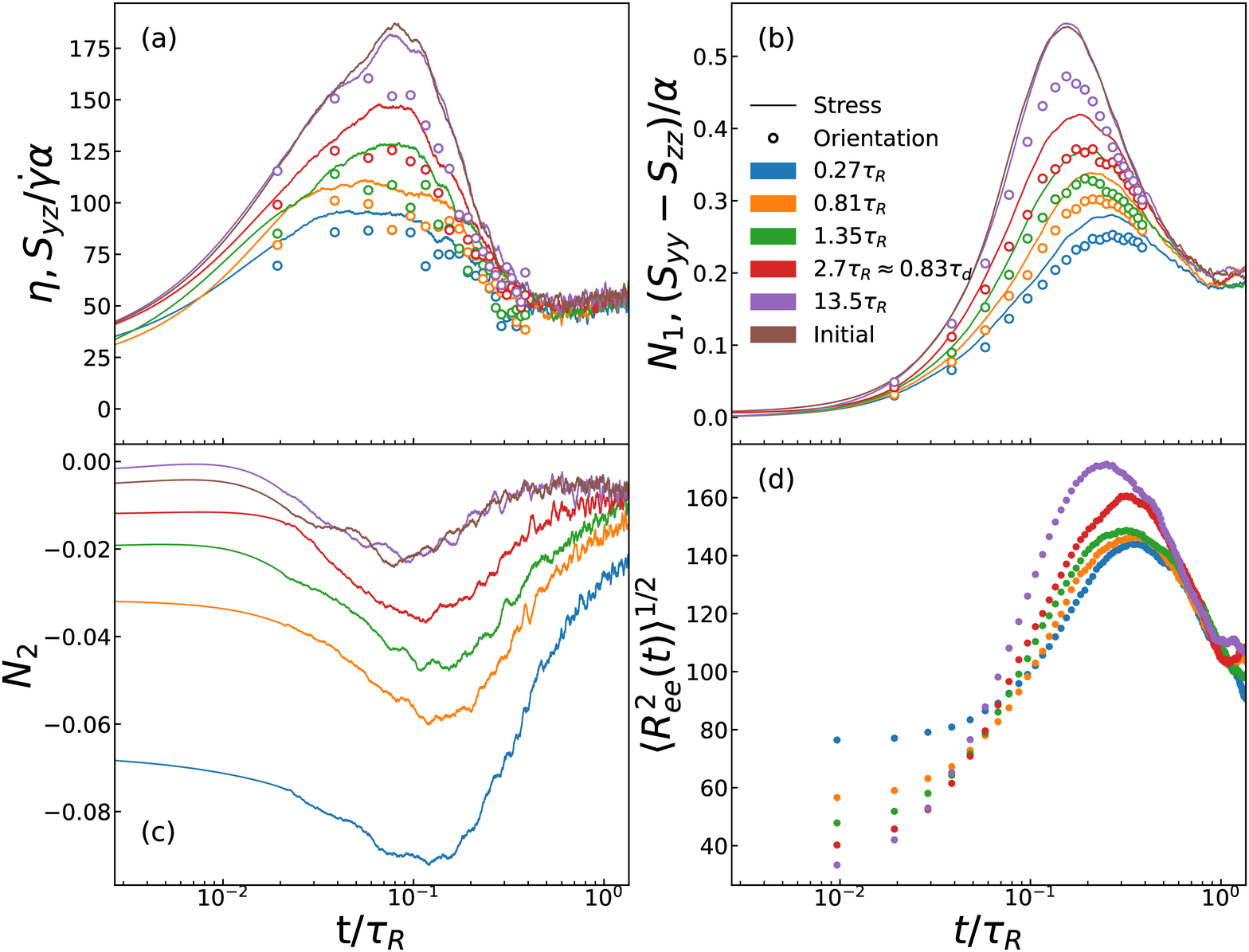}
    \caption[Stress and orientation during orthogonal interrupted shear of an entangled melt.]{Stress components of the fully-flexible entangled melts during orthogonal interrupted shear simulations. Plots show (a) viscosity, (b) first and (c) second normal stress differences and (d) end-to-end distance of chains. The curve labeled initial refers to the first stage of shear, and the other curves refer to the second stage of shear after a waiting time indicated by the label. Open circles in the plots for stress show the components of the bond orientation tensor $\bm{S} = \langle \bm{uu}\rangle$. Parameters: N=500, $Wi_R = 100$.}
    \label{fig:Wi100_long_yz}
\end{figure}

Figure \ref{fig:Wi100_long_yz} shows data equivalent to that shown in Figure \ref{fig:Wi100_long} for the orthogonal case. While the qualitative features are the same, for the same waiting time the relative magnitude of the overshoots in $\eta, N_1, N_2,$ and $R_{ee}$ is larger than in the parallel case. The most striking difference is for short waiting times: in the parallel case the overshoot was almost completely gone, while for orthogonal shear there is still a significant overshoot even for waiting times smaller than the Rouse time, a relative increase of around 50\%. 

Another significant difference can be seen in the values of the second normal stress difference $N_2$ for short waiting, which are significantly more negative than during the initial shear. This is mainly due to our change in the definition of $N_2$ to accommodate the different flow direction: the inclusion of a contribution from $\sigma_{xx}$, which is still not fully relaxed for the smallest waiting times, leads to a more negative initial value of $N_2$.

The increase in the overshoot of the end-to-end distance comes exclusively from the behavior of $R^2_y$ along the new flow direction. The $x$ component, which was the largest one in the first stage of shear, decreases monotonically.

\begin{figure}
    \centering
    \includegraphics[width=\columnwidth]{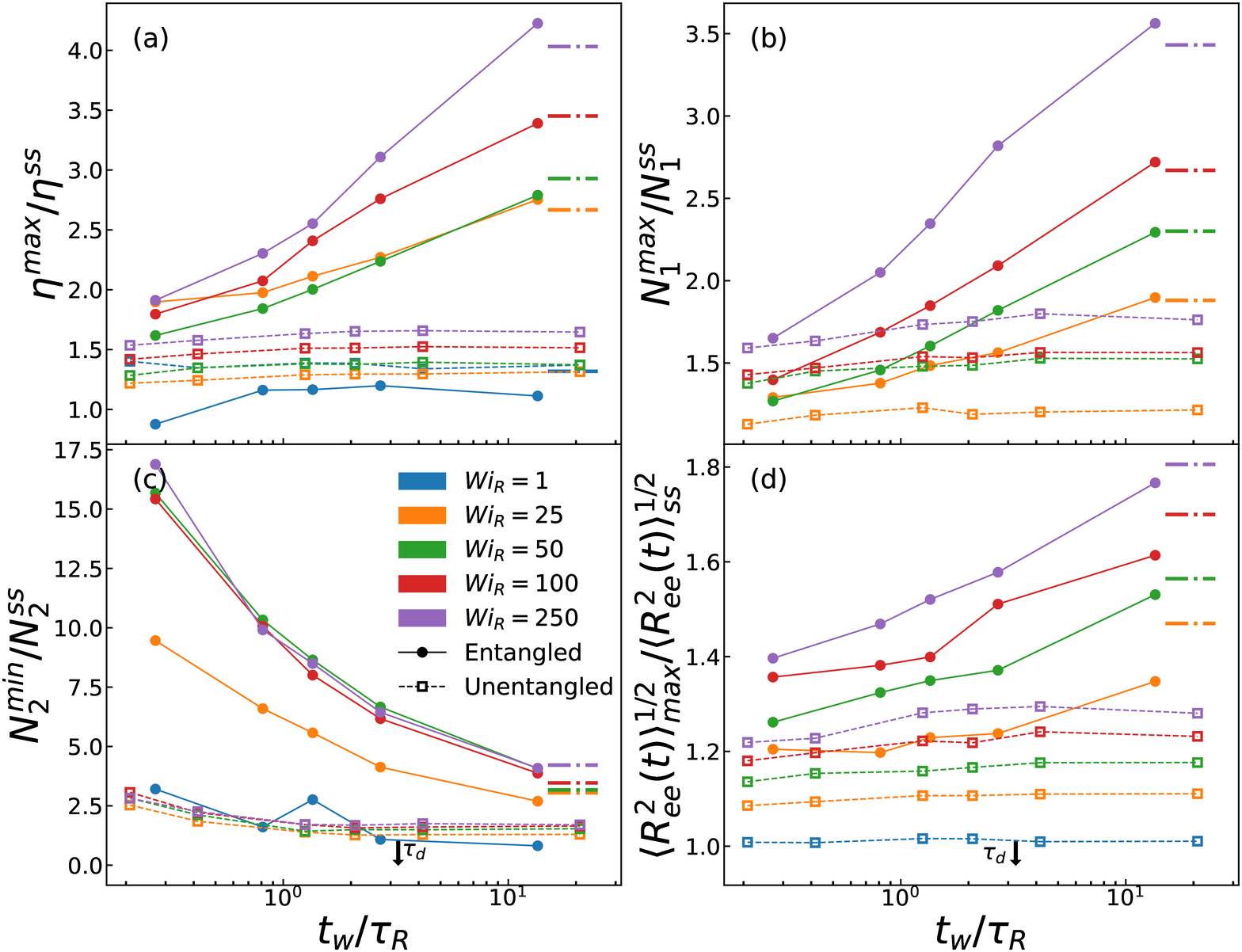}
    \caption[Relative size of overshoots as a function of waiting time for orthogonal interrupted shear.]{Relative size of overshoots and undershoots for entangled (filled circles) and unentangled (open squares) fully-flexible melts during orthogonal interrupted shear as a function of waiting time. Peak of viscosity (a), 1st NSD (b) and end-to-end distance (d) overshoot normalized by steady-state value as a function of waiting time. (c) Minima of the 2nd NSD normalized by steady-state value. The overshoots in the normal stresses for the unentangled case at the lowest Weissenberg number were too small to discern from noise and therefore are not plotted. All properties at the longest waiting times are statistically the same as startup from an equilibrium state. Dash-dotted lines represent the values for the initial startup at the different Weissenberg numbers.}
    \label{fig:max_Wi_comp_yz}
\end{figure}

The relative magnitudes of the overshoots and undershoot are larger than for parallel interrupted shear for all Weissenberg numbers. Figure \ref{fig:max_Wi_comp_yz} shows the values of the overshoot at the peak, normalized by the steady-state values, for both entangled and unentangled systems at all Weissenberg numbers as a function of waiting time. Comparison with Figure \ref{fig:max_Wi_comp} shows that for the shortest waiting times the magnitude of the overshoot is significantly higher when the second stage of shear is orthogonal to the original shear plane.

\begin{figure}
    \centering
    \includegraphics[width=\columnwidth]{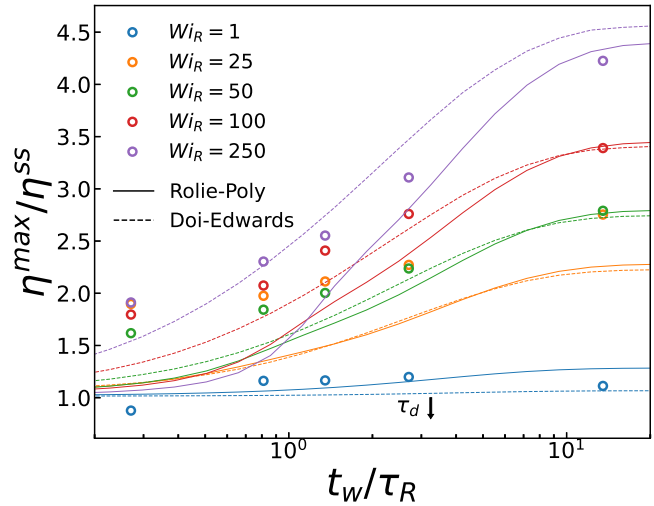}
    \caption[Comparison of constitutive equations to MD data for orthogonal interrupted shear.]{Comparison of Doi-Edwards (dashed) and Rolie-Poly (solid) model predictions for relative size of the viscosity overshoot during the second stage of orthogonal interrupted shear as a function of waiting time to simulation data for the fully-flexible entangled system. Notice the increase in simulation data relative to Figure \ref{fig:data_RP_DE_comp}, while the model predictions remain essentially the same.}
    \label{fig:data_RP_DE_comp_yz}
\end{figure}

Figure \ref{fig:data_RP_DE_comp_yz} compares the viscosity overshoot of the entangled system to predictions by both constitutive models, with the Rolie-Poly parameters $Z = 9$ and $\beta = 1$. There is significant disagreement for the shortest waiting times at all Weissenberg numbers larger than one. In fact, the model predictions are essentially the same as those for the parallel interrupted shear. An analysis of the form of Equations \ref{eq:DE} and \ref{eq:rp} explains this behavior: the coefficients of the relaxation mechanisms in both models involve only scalar quantities such as $\textrm{tr}\bm{\sigma}$, which have no directionality and therefore cannot distinguish stresses related by rotations. The steady-state properties predicted by the models are the same as for parallel interrupted shear.

Figures \ref{fig:data_RP_DE_comp} and \ref{fig:data_RP_DE_comp_yz} together suggest that there is no physical mechanism contained in the analyzed constitutive models that can explain the increase of the relative size of the overshoot due to the change of shear direction. Including extra faster relaxation modes can increase agreement in the short waiting time regime for either parallel or orthogonal interrupted shear, but the same spectrum of relaxation times cannot fit the different values of the overshoot measured in these regimes in the MD simulations. In the next section, we will highlight measurements of chain orientation and entanglements that we believe can help elucidate possible physical mechanisms driving the observed differences in overshoot size. In particular, we propose that the anisotropic nature of the chains in the aligned state leads to remarkably different dynamics of entanglement in the transient regime when flow is restarted in a direction orthogonal to the previous alignment direction, generating a contribution to the stress that arises from the larger number of constraints that need to be overcome in order for the melt to flow into steady state. We argue that constitutive models can be improved upon by utilizing a tensorial definition of entanglements instead of a scalar value of Z.

\section{Entanglements and orientation during parallel and orthogonal interrupted shear}

An important metric in polymer rheology is the number of entanglements per chain. Recent computational studies have focused on how the average number of topological entanglements per chain $\langle Z_k\rangle$ changes under flow in a United Atom model for polyethylene (UA-PE) \cite{baig_flow_2010,nafar_sefiddashti_elucidating_2019,nafar_sefiddashti_evaluation_2017,nafar_sefiddashti_individual_2019,nafar_sefiddashti_steady_2016}, with this number usually measured by the Z1 code. They observe a decrease in the average $\langle Z_k\rangle$ as a function of Weissenberg number in steady state shear, and a small undershoot in $\langle Z_k\rangle$ during the transient under startup shear roughly corresponding to the overshoot in the first normal stress difference. The UA-PE model is much more detailed than  the simpler coarse-grained FENE model used in our simulations, and differences in quantities such as the friction coefficient between chains might change the degree of disentanglement at comparable Weissenberg numbers.

While it is still unclear how to couple the dynamics of entanglements as observed in simulations to constitutive equations, Ianniruberto and Marrucci proposed a model for the entanglement dynamics based on CCR \cite{ianniruberto_convective_2014}, which has been used along with constitutive equations to  model disentanglement during flow \cite{mcilroy_deformation_2017,mcilroy_disentanglement_2017}. A direct comparison of predictions of this equation to the evolution of the number of entanglements in simulation as measured by the Z1 code is hard since a direct relation between the concept of entanglements in tube models and those measured by simulation techniques is still not known\cite{masubuchi_entanglement_2020}. In order to determine how $\langle Z_k\rangle$ is affected by our flow protocol, we analyzed the number of entanglements per chain during both parallel and orthogonal interrupted startup shear as a function of waiting time. 

\begin{figure}
    \centering
    \includegraphics[width=\columnwidth]{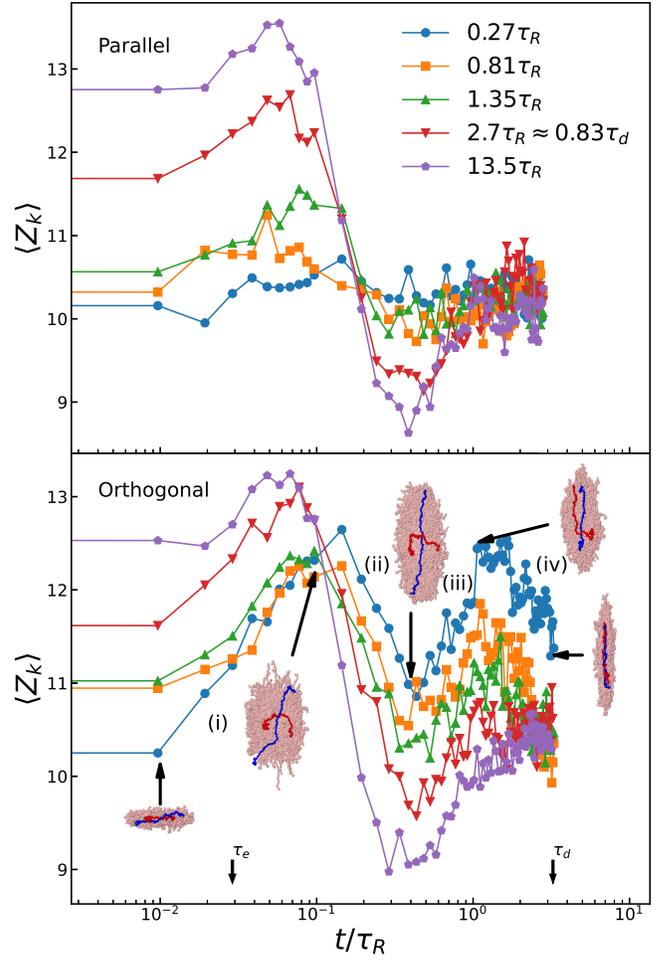}
    \caption{Average number of entanglements per chain, $\langle Z_k\rangle$, as a function of time during the second stage of shear under parallel (top) and orthogonal (bottom) interrupted startup shear. Fully-flexible system, sheared at $Wi_R = 100$. Histograms of $Z_k$ at each snapshot show wide distributions, with standard deviations of order $\pm$ 2-3, corresponding to an error on the mean of approximately $\pm$ 0.1-0.15. Pictures in the bottom plot show all chains in the melt (aligned by their centers of mass), forming a "cloud" that better illustrates the distribution of molecular shapes. Two individual chains in red and blue have been selected for illustration. Chains are visualized in the \textit{xy} plane, with the \textit{x} axis pointing right and \textit{y} axis pointing up. The distinct phases (i), (ii), (iii) and (iv) shown in the bottom panel correspond to the four mechanisms explained in the text.}
    \label{fig:data_Z}
\end{figure}

Entanglement data during interrupted shear for the entangled system are shown in Fig. \ref{fig:data_Z}. During parallel interrupted shear, we observe an undershoot in the number of entanglements per chain at around the same time as the overshoot in the end-to-end distance of the chains. The recovery of the undershoot as a function of waiting time is monotonic, similar to the transient behavior of other measured quantities. We also observe an overshoot during the transient before the undershoot happens, corresponding to a strain similar to the strain at the viscosity overshoot. One possible origin of the brief overshoot is that disentanglement is triggered by the nonaffine deformation of the chains in the nonlinear regime by mechanisms such as CCR, while deformations in the linear regime either preserve the number of entanglements or drive their formation.

During the orthogonal interrupted shear protocol the evolution of the average number of entanglements $\langle Z_k\rangle$ differs remarkably from that observed when shearing from equilibrium. Instead of going through an undershoot into steady-state, $\langle Z_k\rangle$ undershoots and then rises again to a value significantly above the expected steady-state value, finally decreasing on a timescale larger than $\tau_R$. The initial overshoot before the first undershoot is also significantly larger than in the parallel case even for the shortest waiting times. We will demonstrate below that that this interesting behavior can be understood as arising as a superposition of different mechanisms that affect the entanglement structure, and annotate on the bottom panel of figure \ref{fig:data_Z} the regimes where each mechanism dominates the dynamics of entanglements: 
\begin{itemize}
    \item[(i)] the large orientation along an axis orthogonal to the new flow (because the chains have not relaxed after the first shear pulse) enables formation of new entanglements  under shear in a mechanism by which  chains with nearly parallel backbones are pulled over each other; this leads to a large rapid increase in the entanglement number; this can be seen in the bottom panel of figure \ref{fig:data_Z}, where the first and second (from left to right) pictures of the chain "cloud" have similar width, indicating similar alignment along the original flow direction, but the second picture shows a cloud with larger height, indicating chains have started to align along the new flow direction;
    \item[(ii)] disentanglement is triggered by the nonlinear deformation of the chain along the new flow axis; in \ref{fig:data_Z} this can be seen in the difference between the second and third pictures of the "cloud", where the width of the distribution decreases as $\langle Z_k \rangle$ goes down;
    \item[(iii)] reorientation and stretch relaxation in the new vorticity axis lead to reentanglement by forcing chains to collide when aligning along the flow direction; this can be seen in the plot by noticing that the fourth and third pictures of the cloud have a similar aspect ratio despite a large change in $\langle Z_k\rangle$, because most of the changes in the chain orientations are along the axis pointing into the page; 
    \item[(iv)]  in  final stage there is disentanglement to the steady state value after alignment with the new flow axis. Notice how the width of the chain cloud decreases from the fourth to the final picture shown in the plot.
\end{itemize}

This result suggests that topological constraints between chains, as measured by the Z1 code, are inextricably linked to the chain conformation, and a successful model for the dynamics of entanglements should not only describe changes in the average number of entanglements per chain, but also how the created/lost entanglements couple to orientation and flow.

\begin{figure}
    \centering
    \includegraphics[width=\columnwidth]{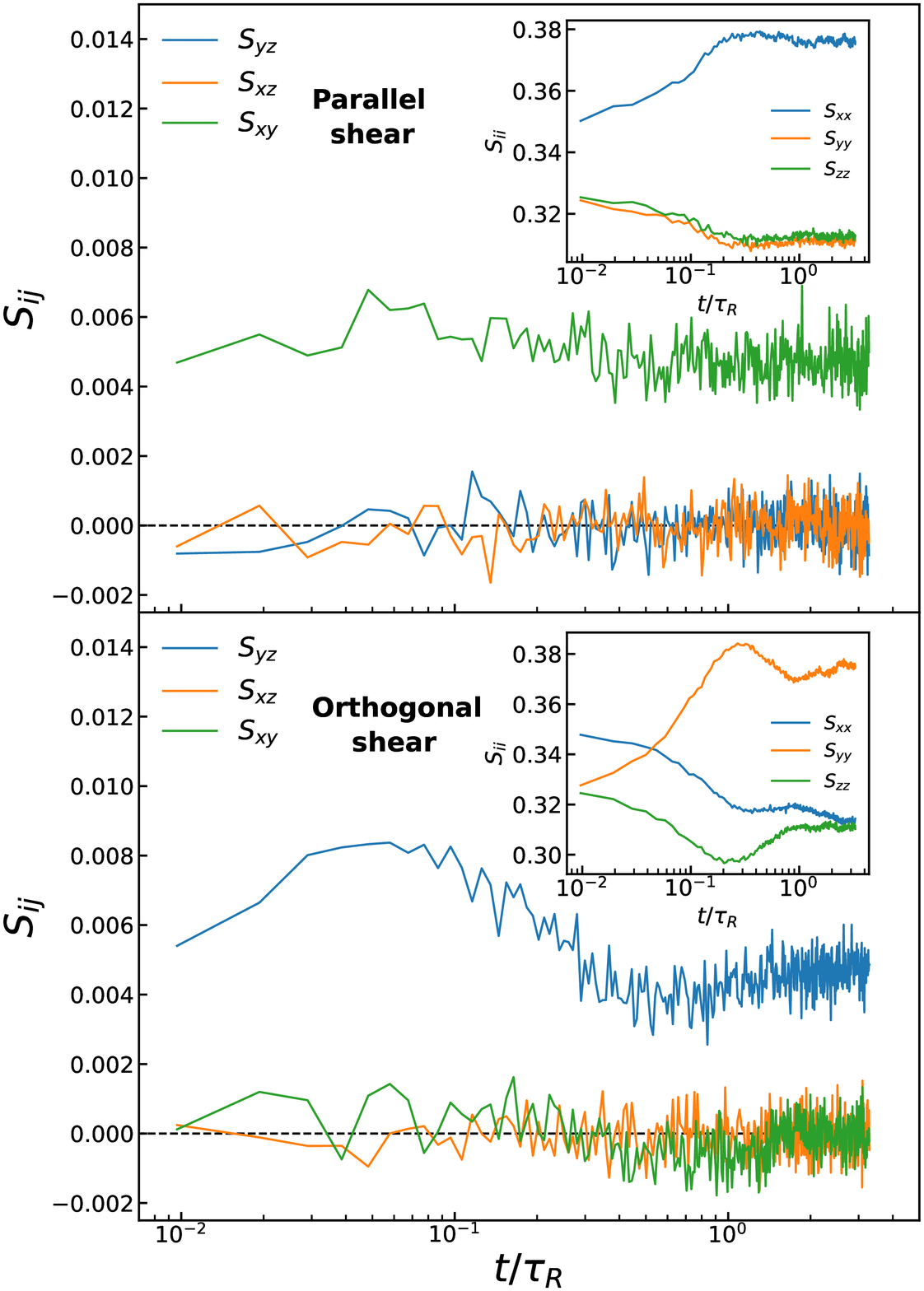}
    \caption{Components of the bond orientation tensor $\bm S$ of fully-flexible entangled ($N=500$) melt in parallel and orthogonal interrupted shear at $Wi_R$ = 100 after a waiting time of $t_w = 0.27\tau_R$. Inset shows diagonal components.}
    \label{fig:data_S}
\end{figure}

\begin{figure}
    \centering
    \includegraphics[width=\columnwidth]{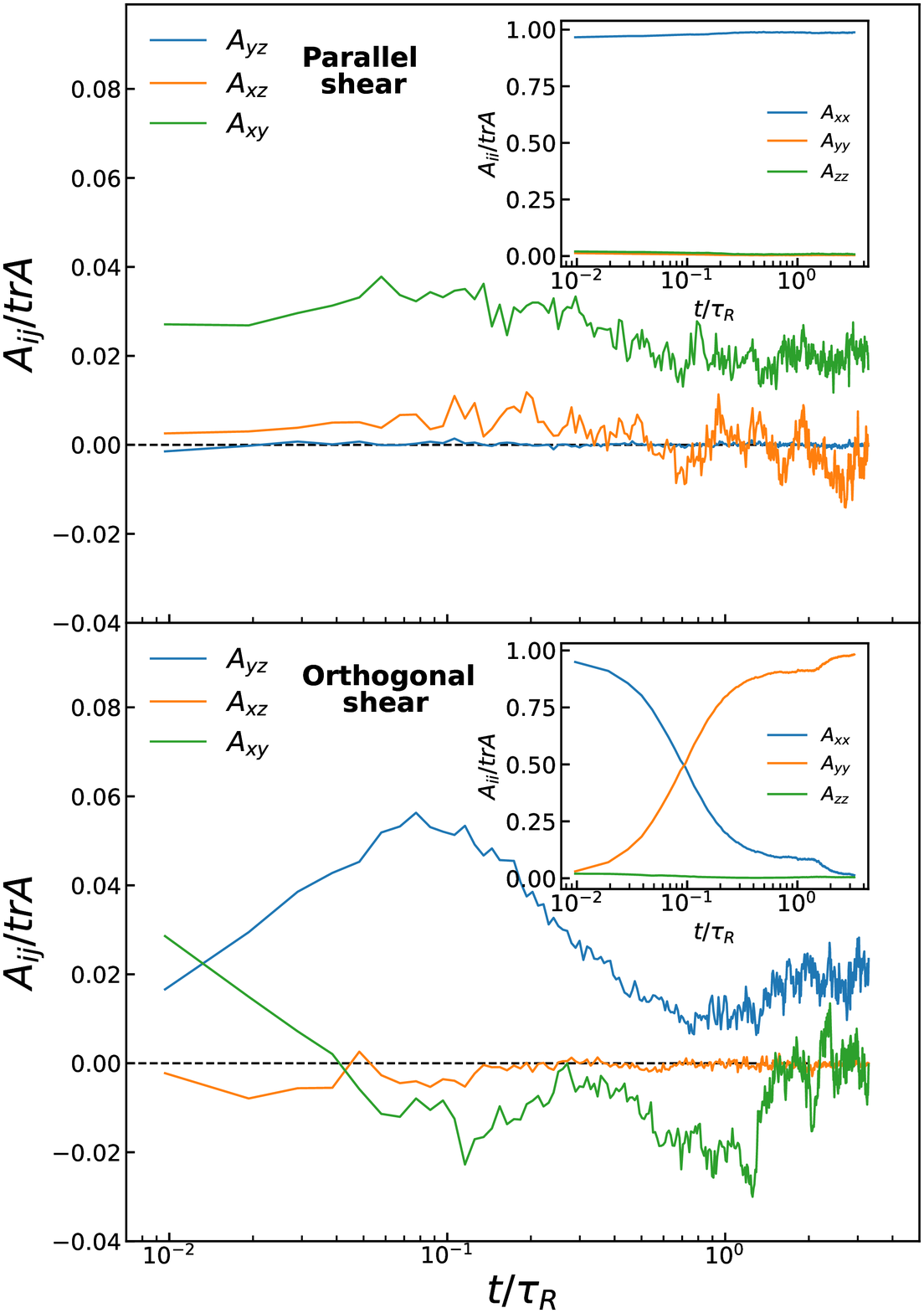}
    \caption{Components of the chain conformation tensor $\bm A$ of fully-flexible  entangled ($N=500$) melt in parallel and orthogonal interrupted shear at $Wi_R = 100$ after a waiting time of $t_w = 0.27\tau_R$. Inset shows diagonal components.}
    \label{fig:data_A}
\end{figure}

One possible way of modelling the evolution of the number of entanglements per chain (or entanglement density) under flow is to assume that the rate of change of the number of entanglements is proportional to the rate of change of the tube length, which can be expressed in terms of components of the bond orientation tensor $\bm S$ and the chain conformation tensor $\bm A = \langle \bm{R}\bm{R}\rangle/2R_g^2$ (where $R_g$ is the radius of gyration and $\bm{R}$ is a chain's end-to-end vector), possibly coupled to the velocity gradient tensor $\bm \kappa$. This approach was used by Ianniruberto and Marrucci \cite{ianniruberto_convective_2014} to obtain an equation for the change in $\nu = Z/Z_{eq}$ under flow which includes a term proportional to $\bm {\kappa : S}$. In order to identify relevant changes in the components of $\bm S$ and $\bm A$ that might be responsible for the curious behavior of the entanglement density under orthogonal interrupted shear, we plot the individual components of these tensors during parallel and orthogonal interrupted shear after our shortest waiting time of $0.27\tau_R$ in Fig. \ref{fig:data_S} and \ref{fig:data_A}. The diagonal components are shown in the insets of the plots and exhibit little change for parallel interrupted shear while showing a reversal of the relevant components for the orthogonal case (relaxing the $xx$ component while increasing the $yy$ component), similar to the behavior observed for the stresses. \par

All components of orientation shown in Figure \ref{fig:data_S} show the same qualitative behavior as the stress curves, as predicted by the stress-optical law. No components under orthogonal shear show the `double peaks' observed for $\langle Z_k\rangle$.

The conformation tensor $\bm A$ combines the orientation and stretch at the \textit{chain} scale into a single measure. The diagonal components of $\bm A$ under flow show very little change in the steady state values of the individual components in parallel interrupted shear but an inversion between $A_{xx}$ and $A_{yy}$ in orthogonal interrupted shear, where $A_{xx}$ relaxes in the same timescale that $A_{yy}$ rises to become the dominant component in $\bm A$. The trace of $\bm A$ is the same as the average end-to-end distance in Fig. \ref{fig:Wi100_long} (d), just squared and normalized by the square of the radius of gyration. \par

An interesting effect of the change in flow direction in the 2nd stage of shear can be seen in the $A_{xy}$ component of the chain conformation tensor during orthogonal interrupted shear. During parallel interrupted shear, both off diagonal components which do not couple to the flow in $\bm{\kappa : S}$ are zero at all times, and one might expect the same would happen to $A_{xy}$ after switching the flow direction. However, our simulations show that $A_{xy}$ goes through a more complicated relaxation process which exhibits a prolonged undershoot, at timescales comparable to the complex behavior appearing in the Z1 data, and featuring two peaks. The shear component $A_{yz}$ exhibits an overshoot similar to the stresses as shown in Fig. \ref{fig:Wi100_long_yz} and is not qualitatively different from what is observed for parallel interrupted shear when taking into account the appropriate flow, shear and vorticity directions (apart from changes in the relative size of the overshoot for the same waiting time, as discussed in Section C). \par

\begin{figure}
    \centering
    \includegraphics[width=\columnwidth]{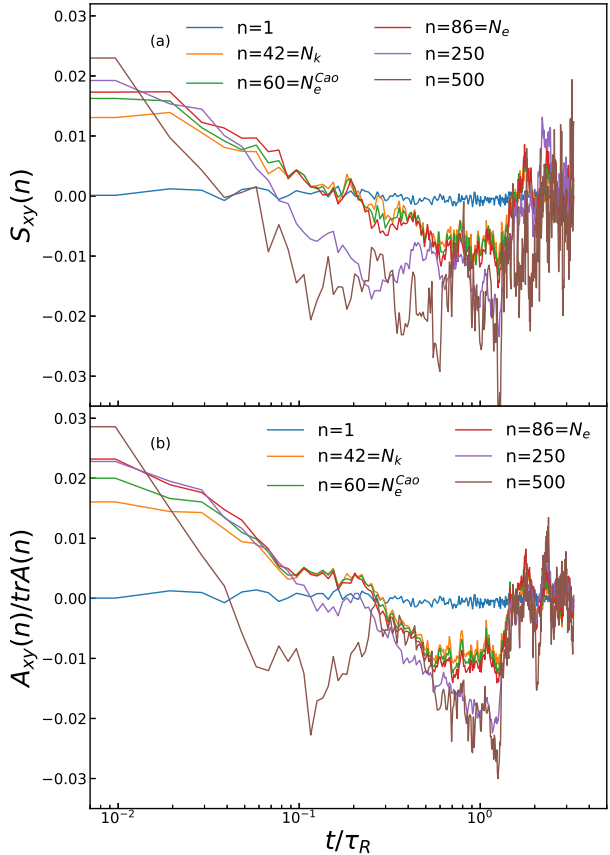}
    \caption{Relaxation of off-diagonal components $S_{xy}$ and $A_{xy}$ of the orientation and conformation tensors under orthogonal interrupted shear at different scales along the contour of the chain. The curves are  $S_{xy}(n) = \langle u_x(n) u_y(n) \rangle$ and ${A_{xy}}(n) = \langle {R}_x(n){R}_y(n)\rangle/2R_{g,n}^2$ where the average is taken over all possible pairs of connecting beads in the same chain separated by a chemical distance $n$ (as described in main text), which is given by the labels for each curve. Data for fully-flexible entangle melt at Weissenberg number $Wi_R = 100$, and the system was allowed to relax for $t_w = 0.27\tau_R$, corresponding to the blue curve in the bottom panel of Fig. \ref{fig:data_Z}. An extended undershoot in orientation and two peaks in the conformation appear at chemical distances larger than the entanglement scale $N_e$.}
    \label{fig:orientation_scales}
\end{figure}

At first glance, the qualitative difference observed between $S_{xy}$ and $A_{xy}$ might seem to come from the contribution of stretch, which is accounted for in $\bm A$ but not in $\bm S$. However, another important difference between the two measurements is the scale over which these measurements are taken. The bond orientation tensor $\bm{S} = \langle \bm{uu}\rangle$ measures orientation at the monomer scale, since $\bm u$ is the bond length between neighboring beads in a chain, while $\bm A$ measures the stretch and orientation of the whole chain. We define orientation and conformation tensors at different scales as $\bm{S}(n) = \langle \bm{u_nu_n}\rangle$ and $\bm{A}(n) = \langle \bm{R}_n\bm{R}_n\rangle/2R_{g,n}^2$, where $R_n$ is the distance vector connecting two beads \textit{n} monomers apart along the chain, $u_n = R_n/|R_n|$ is the normalized distance vectors, $R_{g,n}^2 = R_n^2/6$ is the equivalent radius of gyration and averages are taken over all suitable pairs of monomers. In Figure \ref{fig:orientation_scales}, we show $S_{xy}(n)$ and normalized $A_{xy}(n)$ measured at different relevant scales along the chain during orthogonal interrupted shear in order to understand the contribution of orientation at different scales to the double peak feature observed in $A_{xy}$. We see that while the orientation in the monomer scale does not undershoot the steady state value, the orientation at larger scales does show an undershoot at a timescale close to $\tau_R$, and at scales larger than $N_e$ the prolonged undershoot also appears in $S_{xy}(n)$. Interestingly, the double peak only seems to appear in the conformation tensor, and only for scales comparable to the whole length of the chain. \par

While the relevance of these changes in the $xy$ components of $\bm S$ and $\bm A$ during orthogonal interrupted shear is not clear, a term including off-diagonal components of orientation and conformation that do not couple to flow through terms such as $\kappa : S$ in a kinetic equation for $dZ/dt$ might be able to explain the difference in the dynamics of $Z$ in our two different flow protocols and can be investigated more thoroughly in future work. Terms that contain a relevant contribution from $S_{xy}$ and $A_{xy}$, such as those involving tensor invariants like the determinant, might also need to account for the flow history of the melt at different scales along the chain.

\section{Conclusions}

We used non-equilibrium molecular dynamics simulations to probe the transient rheological properties of polymer melts under startup shear, both starting from the equilibrium configuration and from a state that has relaxed after steady state of shear for a waiting time $t_w$. The observed behavior during startup shear from equilibrium is in line with what has been previously reported in the literature, and shows that the stress optical law holds up to large Weissenberg numbers ($Wi_R >$ 100).

Our simulations of interrupted startup shear show a monotonic recovery of the transient viscosity overshoot with increasing waiting time, as is seen in experiments. Recovery of associated overshoots and undershoots in the normal stresses, end-to-end distance of the chains and number of entanglements per chain are also observed. Constitutive models based on tube theory can qualitatively reproduce the stress measurements, as has been demonstrated extensively before. Semi-quantitative agreement can be obtained by carefully tuning model parameters, though these do not necessarily reflect the parameters as measured in simulations (such as the number of entanglements per chain). Better agreement can probably be obtained if one uses multi-mode versions of the constitutive equations, which we have not used.

\begin{figure}
    \centering
    \includegraphics[width=\columnwidth]{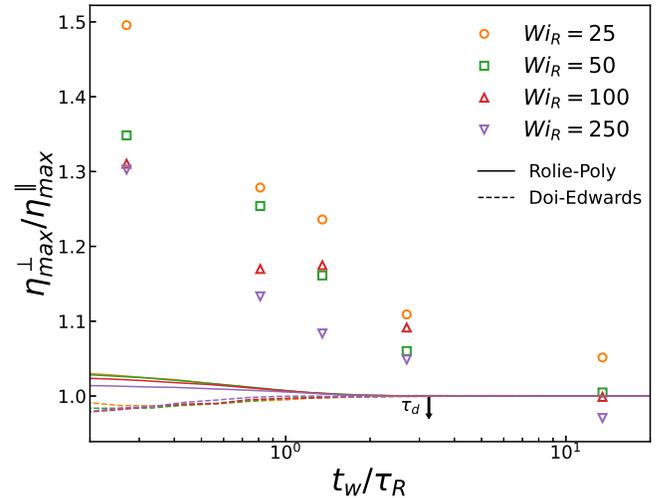}
    \caption[Ratio of peak viscosity under orthogonal interrupted shear to parallel interrupted shear]{Ratio of peak viscosity under orthogonal interrupted shear to parallel interrupted shear in MD simulations of fully-flexible entangled system (open symbols) and predictions by the Doi-Edwards (dashed line) and Rolie-Poly (solid line) constitutive models. The colors used for the MD data are also used to indicate different $Wi_R$ for the model predictions. Data for $Wi_R = 1$ is not shown because data at the peak is too noisy to show a meaningful difference between the orthogonal and parallel cases.}
    \label{fig:ortho_vs_parallel}
\end{figure}

By performing simulations of orthogonal interrupted shear, where shearing is resumed in a different plane after the waiting time, we were able to show that these same models fail to accurately predict the recovery of the overshoot even qualitatively. Both the differential Doi-Edwards model and Rolie-Poly predict the same overshoot size regardless of the relative orientation of the shear flow during the second stage of shear. Figure \ref{fig:ortho_vs_parallel} shows the ratio of the peak value of the viscosity overshoot during orthogonal interrupted shear to parallel interrupted shear in our simulations and in predictions by both constitutive models studied.

The maximum viscosity under orthogonal interrupted shear is up to 50\% larger than the maximum under parallel interrupted shear for the systems studied, while the constitutive models predict at most differences on the order of 3\% for very short waiting times. We note that the maximum viscosity values predicted by our Rolie-Poly model is smaller than the obtained by MD simulations for short waiting times (as can be seen in Figures \ref{fig:data_RP_DE_comp} and \ref{fig:data_RP_DE_comp_yz}) and as a result a 3\% increase is significantly smaller when compared to the 50\% increase seen in MD simulations.
It is easy to see that such behavior can not be modeled by the constitutive models due to the lack of any anisotropic quantities in the incorporated relaxation mechanisms. When integrating the differential equations the only differences between the second and first stage of shear are the relative magnitudes of the coefficients for the different relaxation mechanisms, which depend on time exclusively through scalar quantities. Small differences between the maximum viscosity in orthogonal and parallel interrupted shear can be seen for short waiting times due to the relatively large difference in initial values of the components of the stress tensor. Better models that include the effects of chain orientation or an anisotropic number of entanglements (representing the very anisotropic nature of neighboring chains in aligned states) in the relative magnitudes of the relaxation mechanisms can plausibly solve this issue.

We used the Z1 code \cite{kroger_shortest_2005}, a standard method for analysis of primitive paths of entangled polymers, to determine changes in the number of entanglements per chain during both types of interrupted shear. In parallel interrupted shear, a simple undershoot is observed during the second stage of shear, mirroring the overshoot seen for quantities such as the viscosity, first normal stress difference and end-to-end distance. Increasing the waiting time between stages of shear controls the relative size of the undershoot in $\langle Z_k\rangle$, which changes monotonically with $t_w$. For orthogonal interrupted shear, the same behavior is observed for long waiting times, but short waiting times lead to peculiar behavior of $\langle Z_k\rangle$, which develops two successive overshoots during the second stage of shear. The only components of orientation and conformation that show similar behavior are the relaxing off-diagonal components $S_{xy}$ and $A_{xy}$, which do not couple to flow (since $\kappa_{xy}=0$) in the second stage of shear in standard models of polymer rheology. 

Our work suggests that constitutive models that couple stress to polymer orientation and  entanglements could be modified to accomodate the directional nature inherent in actual entanglement constraints (rather than a smooth scalar quantity), which is evidently discriminated by complex time-dependent flows such as this example of orthogonal interrupted shearing flows. Future work will focus on the effect of chain stiffness on the phenomena described in this paper. In particular, some of our preliminary results show that stress and $\langle Z_k\rangle$ data for melts of stiffer chains still exhibit qualitatively the same features observed here.

\section{Acknowledgements}
Mark O. Robbins sadly passed away during the writing of this manuscript. We thank Vicky Nguyen, Ben Dolata, Jon Seppala, Kalman Migler, Gretar Tryggvason, Martin Kr\"oger, and Thomas O'Connor for advice and discussions. This work was funded by NSF (DMREF 1628974), the Ives Foundation and Georgetown University.
\appendix

\section{Orthogonal interrupted shear with constant gradient direction}

The orthogonal interrupted shear protocol discussed in the main text rotates both the velocity and gradient by ninety degrees:  the gradient direction in the first stage becomes the flow direction in the second stage, while the flow in the first stage becomes the neutral vorticity direction in the second stage. Inverting the flow and gradient direction in the second stage of shear, such that the gradient direction remains the same during both stages of shear and the initial vorticity direction becomes the flow direction during the second stage can lead to a different stress response. This second flow is partially orthogonal, or ``orthogonal-parallel", since the gradient direction remains the same, while the first flow is ``orthogonal-orthogonal".

In order to test this, we repeat the simulations detailed in the main text using a velocity gradient tensor given by
\begin{equation}
    \bm{\kappa} =
    \begin{pmatrix}
      0 & \kappa_{xy} & 0\\
      0 & 0 & 0\\
      0 & \kappa_{zy} & 0
    \end{pmatrix}.
\end{equation}
During the first stage of shear, $\kappa_{xy}=\dot{\gamma}$ and $\kappa_{zy}=0$, while during the second stage $\kappa_{xy}=0$ and $\kappa_{zy}=\dot{\gamma}$. This geometry is similar to that used in experiments by Kraft\cite{kraft_shear_1997,kraft_untersuchungen_1996}.

\begin{figure}
    \centering
    \includegraphics[width=\columnwidth]{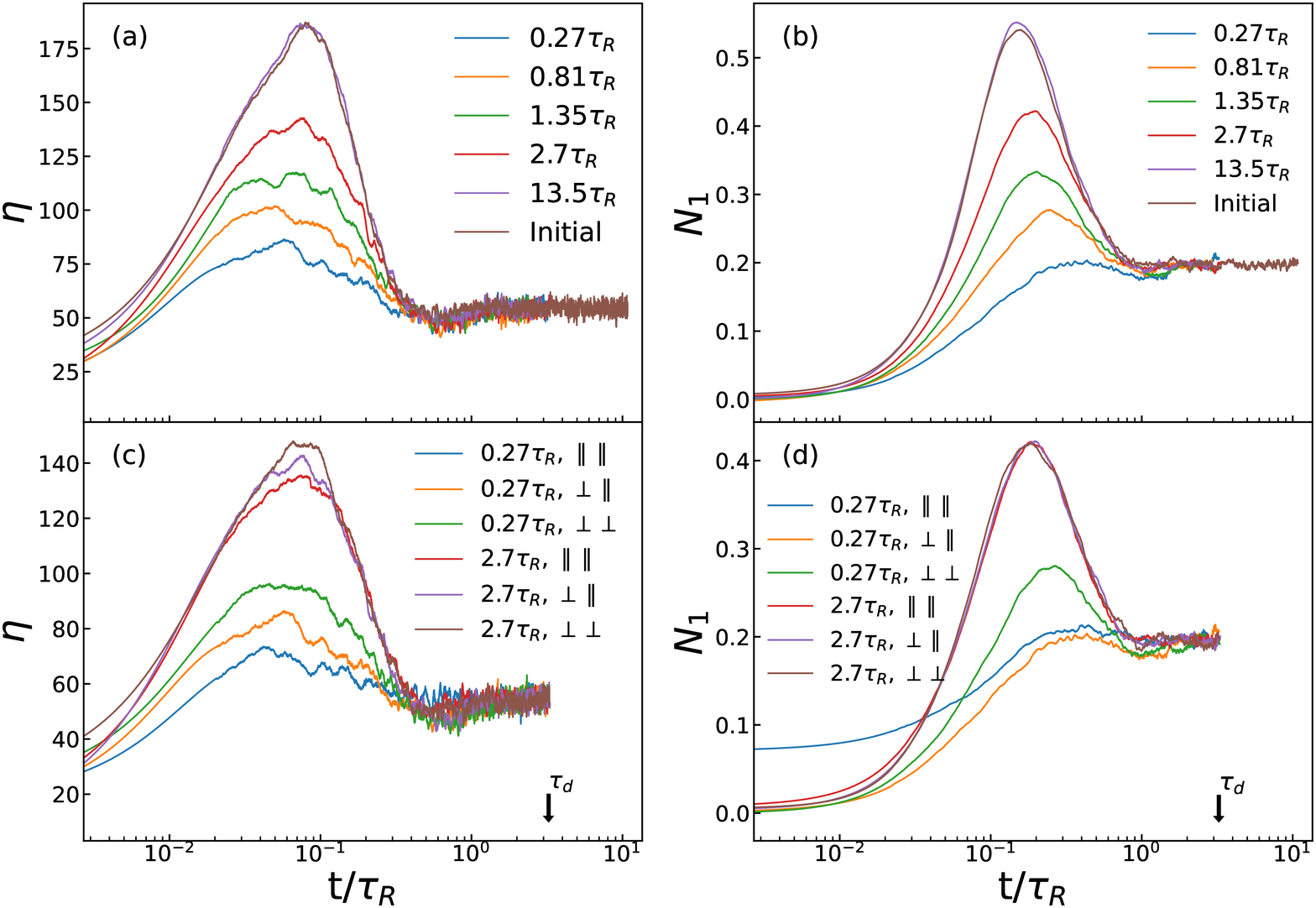}
    \caption{ (a,b) Viscosity and first normal stress differences during orthogonal interrupted shear with constant gradient direction. The monotonic increase of the overshoot is observed for both measures of stress. (c,d) Comparison between viscosity and first normal stress difference curves between our three flow protocols (in legend, $\parallel$ means parallel interrupted shear, $\perp\perp$ means orthogonal-orthogonal interrupted shear ($yz$) and $\perp\parallel$ means orthogonal-parallel interrupted shear ($zy$) with constant gradient) with two different waiting times. For longer waiting times the stress response is similar, but for short waiting times the viscosities are different for the three methods, while the first normal stress difference has a significantly larger peak only for orthogonal shear.}
    \label{fig:stress_comp_xy_yz_zy}
\end{figure}

Figure \ref{fig:stress_comp_xy_yz_zy} shows a comparison between the viscosity and first normal stress difference at $Wi_R = 100$ for the three strain protocols (parallel and the two different kinds of orthogonal) for short and long waiting times.

As expected, there is no difference in the stress response after a long waiting time. However, for a short waiting time the peak value of the viscosity lies between the peaks of parallel shear and the orthogonal-orthogonal shear discussed in the main text, while the peak in the first normal stress difference shows no difference to the one observed for parallel interrupted shear.

\begin{figure}
    \centering
    \includegraphics[width=\columnwidth]{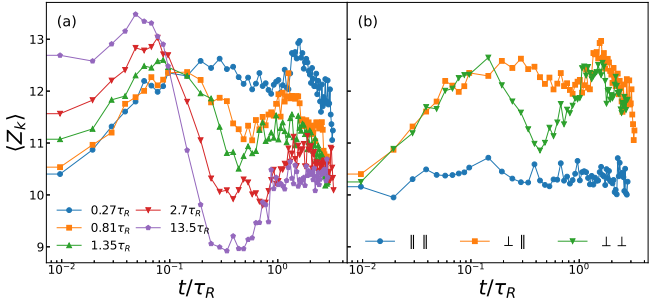}
    \caption{ (a) Evolution of the average number of entanglements per chain $\langle Z_k\rangle$ during orthogonal interrupted shear with constant gradient direction for several Weissenberg numbers. (b) Comparison of the evolution of $\langle Z_k\rangle$ during the three different flow protocols for the shortest waiting time, $t_w = 0.27\tau_R$. Here, $\parallel$ means parallel ($\kappa_{xy}$) interrupted shear, $\perp\perp$ means orthogonal-orthogonal interrupted shear ($\kappa_{yz}$) and $\perp\parallel$ means orthogonal-parallel interrupted shear ($\kappa_{zy}$) with constant gradient. There is almost no change in the parallel case, while in both orthogonal protocols there is a substantial rise in $\langle Z_k\rangle$ during the earlier stages of flow, before a decay for  $t\gtrsim\tau_R$. Additionally, for orthogonal shear $\langle Z_k\rangle$ drops between $\tau_e$ and $\tau_R$, unlike for the other two methods.}
    \label{fig:Z_zy_combined}
\end{figure}

The number of entanglements per chain also displays some similarities to the behavior described in the main text for short waiting times (figure \ref{fig:Z_zy_combined}). The number of entanglements rises as the melt is sheared and then decreases when approaching steady state. Although the initial rise and final decrease in entanglements is very similar to that observed in the fully orthogonal shear protocol,  at short waiting times it lacks the double peak structure found between $0.1\tau_R$ and $\tau_R$ in that case (compare Figures~\ref{fig:data_Z} and \ref{fig:Z_zy_combined}). The characteristic two peaks are recovered for intermediate waiting times, while for the longest waiting times we observed the expected behavior similar to a standard startup shear, with a decrease in $\langle Z_k\rangle$ followed by an undershoot before reaching steady state. The orientation and conformation tensor also lack a clear sign of independent peaks in the off-diagonal components.

\begin{figure}
    \centering
    \includegraphics[width=\columnwidth]{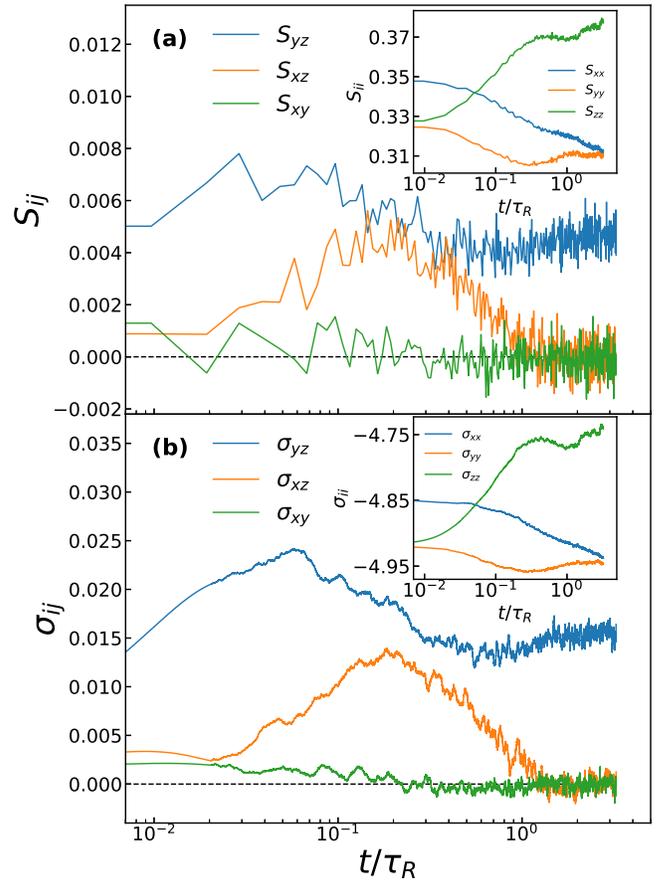}
    \caption{(a) Monomer-scale orientation and (b) stress components during orthogonal interrupted shear with constant gradient direction of the fully-flexible melt after a waiting time of $t_w = 0.27\tau_R$. Components of stress and orientation can be related to one another by the stress-optical law. The overshoot in $S_{xz}$ is reflected in an overshoot in the shear stress $\sigma_{xz}$.}
    \label{fig:orientation_zy}
\end{figure}

The orientation evolution during orthogonal-parallel interrupted shear also shows striking differences at short waiting times, compared to that observed for the orthogonal-orthogonal shear (see Fig.~\ref{fig:orientation_zy}). The $\sigma_{xz}$ component of the stress tensor, which is not the primary imposed stress during either the first or second stages of shear (where $xy$ and $zy$ are respectively imposed) has a pronounced overshoot in the orthogonal-parallel case which is not seen in the other flow protocols. Since the $xz$ component couples the flow directions during the first ($x$) and second  ($z$) stages of shear, we speculate that this overshoot is due to the realignment of chains when the flow direction is changed, leading to a noticeable bump in stress. This is similar to the small undershoot observed during startup that we associate with the tumbling of chains. 

As described by the stress-optical law (Eq. \ref{eq:sor}), this change in orientation leads to a change in the corresponding component of stress. This change in $\sigma_{xz}$ is shown in Figure \ref{fig:orientation_zy}. As a general rule, all flow protocols satisfy the stress-optical law for the entire orientation and stress tensors, with systematic deviations at higher Weissenberg numbers (see Figure \ref{fig:startup}) and for unentangled melts. Kraft \cite{kraft_shear_1997,kraft_untersuchungen_1996} performed experiments with a similar setup and an effective waiting time $t_w = 0$. However, they only monitored $\sigma_{xy}$ and $\sigma_{zy}$, observing behavior similar to the one we see in the simulations, and did not show measurements of $\sigma_{xz}$. Experimental measurement of a non-zero shear stress in the $xz$ plane during orthogonal interrupted shear with constant gradient direction could then be another way of verifying the results of our simulations for model linear polymer melts.

\label{sec:app_a}

\section{Changes in average number of entanglements for different Weissenberg numbers}

In the main text we showed changes in the average number of entanglements per chain $\langle Z_k\rangle$, as measured by the Z1-code, as well as in the components of orientation and conformation tensors during parallel and orthogonal interrupted shear simulations of our entangled fully flexible melt at $Wi_R = 100$. For completeness, in this appendix we show the changes in the same quantity for different Weissenberg numbers.

\begin{figure*}
    \centering
    \includegraphics[width=\textwidth]{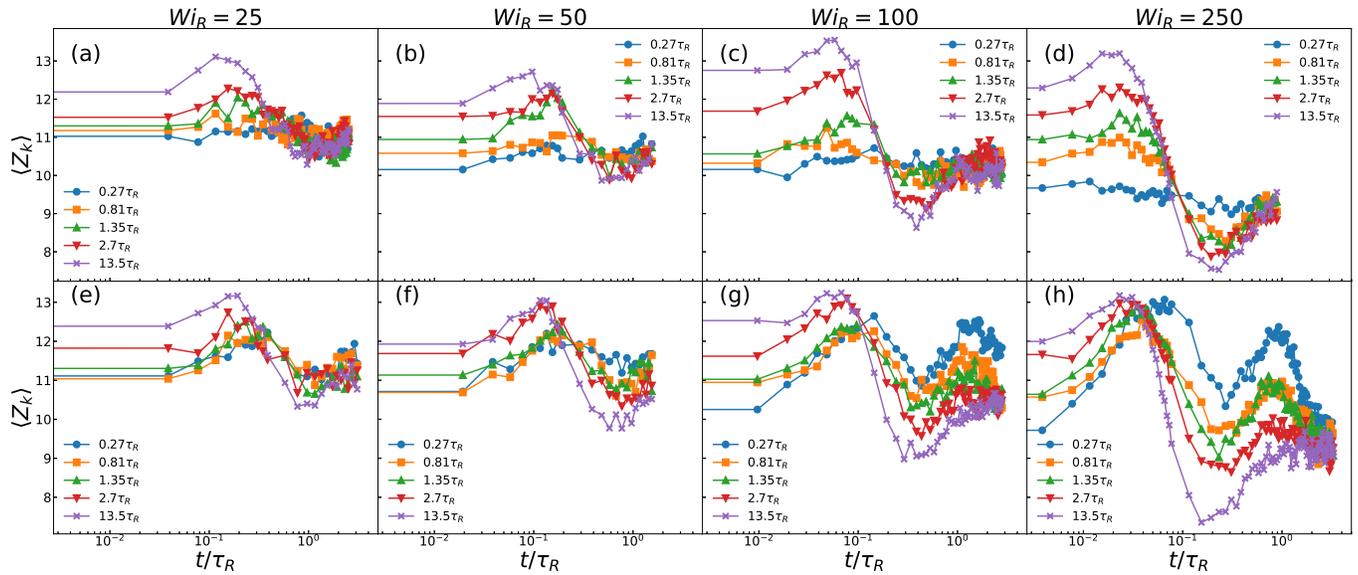}
    \caption{(a-d) Changes in the average number of entanglements per chain $\langle Z_k\rangle$ during parallel interrupted shear for Weissenberg numbers 25, 50, 100 and 250, from left to right. The different colors represent different waiting times. (e-h) Same data for orthogonal interrupted shear. Note the strong deviations from parallel behavior for short waiting times, and how these features are enhanced by increasing the relative flow strength.}
    \label{fig:Z_Wis}
\end{figure*}

Figure \ref{fig:Z_Wis} shows data for $\langle Z_k\rangle$ under parallel and orthogonal interrupted shear of our fully flexible melt for Weissenberg numbers from 25 to 250. While data is available for $Wi_R = 1$, the absolute change in the value of $\langle Z_k\rangle$ is very small because to the weak imposed flow.

\label{sec:app_b}

\nocite{*}
\bibliography{intshear}

\end{document}